\documentclass[12pt]{article} 
\pdfoutput=1
\usepackage{amssymb,graphicx}
\usepackage{epstopdf}
\usepackage{amsmath,amsfonts}
\usepackage{epsfig} 
\usepackage{graphicx,graphics}
\usepackage{xcolor}
\usepackage[unicode=true,pdfusetitle,
bookmarks=true,bookmarksnumbered=true,bookmarksopen=true,bookmarksopenlevel=3,
 breaklinks=false,pdfborder={0 0 1},backref=false,colorlinks=true,citecolor=blue]
 {hyperref}

\begin{document} 

\title{\bf Caustic formation in DBI models:\\ 
Wave propagation on planar domain walls}
\author{E.~Babichev$^a$, B.~Gafarov$^b$, S.~Ramazanov$^c$, M.~Valencia-Villegas$^{c,d}$ \\
\small{\em $^a$Universit\'e Paris-Saclay, CNRS/IN2P3, IJCLab, 91405 Orsay, France}\\
\small{\em $^b$Faculty of Physics, MSU, 119991 Moscow, Russia}\\
\small{\em  $^c$Institute for Theoretical and Mathematical Physics, MSU, 119991 Moscow, Russia}\\
\small{\em $^d$Institute for Nuclear Research of the Russian Academy of Sciences, 117312 Moscow, Russia}\\
}
 
 \date{}

{\let\newpage\relax\maketitle}

\begin{abstract}
We investigate propagation of \textit{generic waves} on thin planar domain walls effectively described by the scalar Dirac-Born-Infeld model (DBI). We pay a particular 
attention to the possibility of caustic formation --- the process, which may lead to intensive particle emission by domain walls. It is demonstrated that no singularities arise in DBI 
in 2D flat spacetime in the hyperbolic case, if one starts from smooth initial conditions. Technically, this happens because the same family characteristics of the relevant partial differential equation remain parallel at all the times, albeit not 
being straight lines generically. Crucially, characteristic curves cease to be parallel beyond the simplified setup of DBI in 2D flat spacetime. 
In particular, this is shown to be the case in $D>2$ for spherical waves, in an expanding Universe, and in the case of a minimal deformation of DBI necessary for avoiding the domain wall problem in cosmology. However, we prove that DBI remains caustic free in the hyperbolic case in all these physically relevant situations.  This strongly suggests that caustics can form on planar domain walls only due to the loss of hyperbolicity, and they have a cusp profile. We demonstrate, how the non-trivial structure of DBI characteristics beyond the 2D flat spacetime setup uncovered in this work can significantly affect cusp formation.
\end{abstract}

\section{Introduction}

Caustic formation at intersections of particle trajectories is a widespread phenomenon emerging in various areas of physics. The appearance of caustics manifests breakdown of an effective field theory (EFT) 
approach, which must be replaced by a more fundamental one, and thus they are of considerable theoretical and phenomenological interest. Namely, certain physical quantities become multi-valued 
at the particle intersections in the EFT framework, while their derivatives turn into infinities. Caustics famously appear in geometric optics, which is a zero wavelength limit of classical electrodynamics~\cite{Orlov}. Furthermore, a pressureless perfect fluid made of collisionless dust particles is vulnerable to caustics~\cite{Zeldovich:1969sb, Arnold}. In the realistic theory involving particle interactions, singularities are regularized by non-zero scattering cross-sections, while caustics signal the onset of multistream regime.

 In this work we search for the possibility of caustic formation in Dirac--Born--Infeld model (DBI)~\cite{Born:1934gh, Dirac:1962iy}, more precisely its scalar version~\cite{Barbashov:1966frq, Barbashov:1966nvq}. The scalar DBI is described by the Lagrangian\footnote{Hereafter we focus exclusively on subluminal DBI.} ${\cal L}=-\sqrt{1-2X}$, where $X$ is the canonical kinetic term of 
 the field $\phi$. Such a construction is directly related to the Nambu-Goto action, which appears in different physics contexts, e.g., in string theory~\cite{Tseytlin:1999dj} and in cosmology. For example, 
  DBI describes cosmic domain walls~\cite{Zeldovich:1974uw, Vilenkin, Vachaspati} in the limit of infinitely small wall width\footnote{See, e.g., Refs.~\cite{Frolov:2002rr, Alishahiha:2004eh} for other, string theory inspired, manifestations of DBI in cosmology.}, see Sec.~\ref{todbi}. 
Understanding particularities of domain wall evolution is the primary motivation of this work. Note, however, that our results 
 are applicable to any model effectively described by DBI, unless domain walls are mentioned explicitly.

Naturally, formation of caustics is a violent high energy phenomenon. Thus, it is reasonable to consider DBI caustics as sources of heavy particles, --- quanta of the field constituting domain walls. In turn, 
particle emission may play the instrumental role for settling walls to the scaling regime, which has been confirmed numerically~\cite{Press:1989yh, Hiramatsu:2013qaa, Dankovsky:2025pjg, Blasi:2025tmn}. That is, the domain wall network evolves in a self-similar manner, so that there is on average one long wall stretching throughout the observable Universe at any time after reaching the scaling. The wall stays sufficiently smooth, and its curvature radius is maintained to be of the order of the particle horizon. Gravitational wave emission by domain walls~\cite{Hiramatsu:2013qaa, Dankovsky:2024zvs} is insufficient to keep them smooth, and thus particle emission is likely to be the only essential mean of energy release. This discussion of scaling is akin to that in the case of cosmic strings~\cite{Kibble:1976sj}, where such phenomena as cusps and kinks are known to be responsible for the gravitational wave and particle production~\cite{Vilenkin, Vachaspati:1984gt, Vilenkin:1986ku, Blanco-Pillado:2017oxo, Baeza-Ballesteros:2024sny}. Caustics in DBI have been previously reported in Refs.~\cite{Felder:2002sv, Eggers:2009zz, EHS, Blanco-Pillado:2025gzs}, 
    and they are indeed analogous to cosmic string cusps, at least in 2D~\cite{Blanco-Pillado:2025gzs}. These caustics referred to as cusps also in the domain wall context, are linked to the loss of hyperbolicity in DBI, and we comment on them in the present work.

  The question of our primary interest, however, is the possibility of caustic formation in DBI, provided that the hyperbolic condition is fulfilled. The strategy for the search of caustics in generic $P(X)$-theories (to which DBI belongs) in the hyperbolic case has been elaborated in Ref.~\cite{Babichev:2016hys}. The latter demonstrates that $P(X)$-theories are vulnerable to caustic singularities using the tractable example of simple waves in 2D. DBI is exceptional, though. In particular, Ref.~\cite{Mukohyama:2016ipl} uses the techniques of Ref.~\cite{Babichev:2016hys}, which we review in Sec.~\ref{method}, to prove that 
DBI simple waves propagate smoothly in 2D flat spacetime. The earlier discussion on the subject can be found in Ref.~\cite{Deser:1998wv}, which in turn builds upon Ref.~\cite{Boillat:1970gw} arguing for the exceptional status of the electromagnetic version of DBI, cf. Ref.~\cite{Blokhintsev:1953kip}. 

We revisit DBI in 2D flat spacetime and confirm that it is caustic free (in the hyperbolic case), see Secs.~\ref{dbicharacteristics} and~\ref{no}. Throughout this work 
we assume smooth initial conditions for the DBI field. This is in contrast to Refs.~\cite{Deser:1998wv, Tanahashi:2017kgn} allowing for discontinuities in the initial data. Compared to Ref.~\cite{Mukohyama:2016ipl}, we prove that not only simple waves but also generic waves propagate without developing caustics. This is accomplished by identifying characteristic curves, which can be interpreted as trajectories of high energy particles. We observe that characteristics belonging to the same family remain parallel, though not being straight lines generally, see Fig.~\ref{no_caustics}. Hence, they do not cross, and caustics do not form.

The situation is less straightforward beyond DBI in 2D flat spacetime. We demonstrate that characteristics cease to be parallel in a row of physically relevant situations, i.e., for spherical waves in the spacetime with $D>2$, in the expanding Universe, and if one allows for certain departures from DBI. Nevertheless, we prove that (modified) DBI remains 
caustic free, if hyperbolicity is fulfilled, see Sec.~\ref{beyond}. Namely, parallelism of characteristics is being restored exactly where one would expect them 
to cross. This can be interpreted as a repulsion force, which gets stronger as characteristic 
curves come close to each other. We conclude that the only possible DBI caustics are likely to be linked to the loss 
of hyperbolicity, and they have a cusp profile, see Sec.~\ref{cuspy}. While the characteristic curves do not cross 
in the hyperbolic case, their non-trivial pattern uncovered in Sec.~\ref{beyond} is expected to have a strong impact on cusp formation, as demonstrated in Figs.~\ref{e_DBI_cusp} and~\ref{H_DBI}.

The outline of the paper is as follows. In Sec.~\ref{todbi} we discuss how DBI arises in the context of domain walls. 
In Sec.~\ref{method} we review the method of characteristics suitable for solving equations of motion in generic $P(X)$-theories. 
We apply this method to DBI in Sec.~\ref{dbicharacteristics}. Basing on results of Sec.~\ref{dbicharacteristics}, we prove the absence of caustics in DBI in 2D Minkowski space in the hyperbolic case in Sec.~\ref{no}. We demonstrate that the caustic free nature of DBI persists beyond this simple scenario in Sec.~\ref{beyond}, despite that characteristics cease to be parallel. In Sec.~\ref{cuspy} we discuss DBI caustics associated with the loss of hyperbolicity. In the discussion section~\ref{discussions} we comment on the relevance of caustics for domain wall evolution.

\section{From domain walls to DBI}
\label{todbi}

 As has been mentioned above, DBI provides an effective description of domain walls emerging in the early Universe when discrete symmetries get spontaneously broken~\cite{Zeldovich:1974uw}. 
The simplest scenario giving rise to domain walls involves a scalar field $\Psi$ endowed with $Z_2$-symmetry spontaneously broken by a non-zero expectation value $\langle \Psi \rangle$: 
\begin{equation}
\label{dw}
S_{DW}=\int d^4 x \sqrt{-g} \left[-\frac{1}{2} g^{\mu \nu} \partial_{\mu} \Psi \partial_{\nu} \Psi-\frac{\lambda}{4} (\Psi^2 -\langle \Psi \rangle^2)^2 \right] \; ,
\end{equation}
where $\lambda$ is the self-interaction coupling constant. (The mostly plus signature of the spacetime metric is assumed hereafter.) In what follows, we consider the domain wall width estimated as $\delta_{wall} \simeq \sqrt{2}/(\sqrt{\lambda} \langle \Psi \rangle)$ to be small compared to other relevant scales, which are typically of the order of the inverse Hubble rate in the cosmological setup. In this thin wall approach, the domain wall is described by the Nambu-Goto action~\cite{Vilenkin}\footnote{See Refs.~\cite{Gregory:1989gg, Bonjour:2000ca, Blanco-Pillado:2024bev} discussing finite wall width corrections to this action.}:
\begin{equation}
\label{NG}
S_{NG}=- \sigma \int d^3 \zeta \sqrt{|\gamma |} \; .
\end{equation}
Here $\sigma =2\sqrt{2\lambda} \langle \Psi \rangle^3/3$ is the wall tension, and $\gamma$ is the determinant of the induced metric $\gamma_{ab}$ on the wall. The metric $\gamma_{ab}$ is given by 
 \begin{equation}
 \label{induced}
 \gamma_{ab}=g_{\mu \nu} \frac{\partial x^{\mu}}{\partial \zeta^{a}}\frac{\partial x^{\nu}}{\partial \zeta^{b}} \; ,
 \end{equation}
 where $\zeta^{a}$ are the coordinates on the worldsheet. In the expanding Universe the spacetime interval is described by $ds^2=a^2 (\eta) (-d\eta^2+dx^2_i)$, where $a (\eta)$ is the scale factor and $\eta$ is the conformal time related to the cosmic time $t$ by $\eta=\int dt/a(t)$. In Appendix A, we discuss the Nambu-Goto dynamics in the conformal gauge commonly employed when studying cosmic strings~\cite{Vilenkin}. In the main text, we assume the static gauge corresponding to the choice of coordinates on the wall $\zeta^{0}=\eta$, $\zeta^{1}=x$, $\zeta^{2}=y$. In that case, one has $\gamma=-a^6 (1-2X)$, where 
 \begin{equation}
 \label{X}
 X=-\frac{1}{2} \eta^{a b} \partial_{a} z \partial_{b} z, 
 \end{equation}
and $\eta^{ab}$ is the metric of the 3D flat spacetime. We end up with the DBI action: 
\begin{equation}
\label{dbireal}
S_{DBI}=-\sigma \int d\eta d x dy a^3 (\eta) \sqrt{1-2X} \; .
\end{equation}
In what follows we identify the spacetime coordinate describing the position on a wall with a scalar field $\phi$: 
\begin{equation}
z \equiv \phi \; .
\end{equation}
It is important here that 
 DBI is suitable for the description of waves on a planar domain wall, so that the field $\phi$ remains single valued. Therefore, breakdown of DBI does not necessarily invalidate the thin wall approach, but may point to the fact that the field $\phi$ tends to become multi-valued, i.e., closed domain walls are about to form.

As it follows, planar domain walls in 4D spacetime are described by the DBI action in 3D. 
To keep the things tractable, however, one suppresses dependence on one spatial coordinate. 
In this case, domain walls are effectively living in 3D spacetime. They are manifested as domain wall strings and described by the 2D DBI action. We will also switch off the cosmic expansion, so that the DBI action simplifies to  
\begin{equation}
\label{dbi}
S_{DBI}=-\sigma \int dt d x \sqrt{1-2X} \; .
\end{equation}
We assume this action when discussing DBI in the next two sections. We will return to a more realistic 
case~\eqref{dbireal} in Sec.~\ref{beyond} and observe significant differences compared to the simplified case~\eqref{dbi}.

\section{Method of characteristics}
\label{method}

Our goal is to solve the partial differential equation (PDE) following from the DBI action by the method of characteristics~\cite{Courant1, Courant2, Vladimir}. 
In this section, we discuss application of this method to generic 
$P(X)$-theories described by the action $S=\int d^4 x \sqrt{-g} P(X)$, closely following Ref.~\cite{Babichev:2016hys}. Here $P(X)$ is a generic function of a scalar field $\phi$ kinetic term. We specify to the case $P(X)=-\sigma \sqrt{1-2X}$ corresponding to DBI in the next section. The scalar field $\phi$ is assumed to live in 2D flat spacetime described by the coordinates $t, x$. 
It is convenient to introduce the notations 
\begin{equation}
\tau =\dot{\phi} \qquad \chi=\phi' \; ,
\end{equation}
where the dot and the prime stand for the derivatives with respect to time $t$ and spatial coordinate $x$, respectively. In terms of $\tau$ and $\chi$, the canonical kinetic term $X$ can be written as $X=\frac{1}{2} (\tau^2-\chi^2)$, while the equation of motion in the generic $P(X)$-theory takes the form
\begin{equation}
\label{partial}
A \dot{\tau}+2B \tau'+C\chi'=0 \; ,
\end{equation}
where
\begin{equation}
\label{ABC}
A =P_X+\tau^2 P_{XX} \; , \qquad B=-\tau \chi P_{XX} \; , \qquad  C=-P_X +\chi^2 P_{XX} \; .
\end{equation}
Here $P_X$ and $P_{XX}$ denote the first and second derivative with respect to $X$, correspondingly. We have used the consistency condition $\tau'=\dot{\chi}$ in Eq.~\eqref{partial}.

The strategy for solving PDE~\eqref{partial} by the method of characteristics is as follows. One introduces characteristic curves parametrized by a variable $\omega$ and described by a slope $\xi=(\partial x/\partial \omega)/(\partial t/\partial \omega)$, such that along these curves the PDE can be recast into the set of ordinary differential equations (ODEs). Along characteristic lines we have 
\begin{equation}
\label{interme}
\frac{\partial \tau}{\partial \omega}= \tau' \frac{\partial x}{\partial \omega}+\dot{\tau} \frac{\partial t}{\partial \omega} \qquad \frac{\partial \chi}{\partial \omega}= \chi' \frac{\partial x}{\partial \omega}+\tau' \frac{\partial t}{\partial \omega} \; .
\end{equation}
(We again took into account that $\tau'=\dot{\chi}$.) We resolve Eq.~\eqref{interme} with respect to $\dot{\tau}$ and $\chi'$, and substitute the corresponding expressions into Eq.~\eqref{partial}. Enforcing that the term $\sim \tau'$ vanishes in the resulting equation, one obtains the constraint on the slope $\xi$ of a characteristic curve:
\begin{equation}
\label{characteristic}
A\xi^2 -2B\xi +C=0 \; .
\end{equation}
This quadratic equation yields two solutions:  
\begin{equation}
\label{determinant}
\xi_{\pm}=\frac{B \pm \sqrt{B^2-AC}}{A} \; ,
\end{equation}
provided that $B^2-4AC>0$, or equivalently if Eq.~\eqref{partial} is hyperbolic. We postpone the discussion of the non-hyperbolic case, when $\xi_{+}=\xi_{-}$, till Sec.~\ref{cuspy}.
As it follows, there are two branches of characteristic curves simply referred to as $\xi_{+}$ and $\xi_{-}$-characteristics in what follows. Respectively, there are two types of parameters: $\omega_{+}$ and $\omega_{-}$. The first pair of ordinary differential equations equivalent to PDE~\eqref{partial} defines characteristics in the $(t, x)$ plane for given $\xi_{\pm}$:
\begin{equation}
\label{char}
\frac{\partial x}{\partial \omega_{+}} = \xi_{+} \frac{\partial t}{\partial \omega_{+}} \qquad \frac{\partial x}{\partial \omega_{-}} = \xi_{-} \frac{\partial t}{\partial \omega_{-}}\; .
\end{equation}
The second pair describes evolution of the field derivatives $\tau$ and $\chi$ along characteristics:
\begin{equation}
\label{ode}
 \frac{\partial \tau}{ \partial \omega_{+}}  + \xi_{-}  \frac{\partial \chi}{\partial  \omega_{+}} =0 \qquad   \frac{\partial \tau}{\partial \omega_{-}}   + \xi_{+}  \frac{\partial \chi}{\partial \omega_{-}}=0\; .
\end{equation}
So, along $\xi_{+}$-characteristics $\omega_{+}$ is varying, while $\omega_{-}=\mbox{const}$, and vice versa for $\xi_{-}$-characteristics. All in all, one can use the parameters $\omega_{+}$ and $\omega_{-}$ as a set of new coordinates when describing dynamics of the field $\phi$. These are the natural coordinates from the following point of view: characteristics describe propagation of high energy particles, and their slopes $\xi_{\pm}$ can be interpreted as phase velocities. 

The latter motivates an equivalent definition of characteristics, which will be useful in the next sections. Let us consider the total differential of $t(\omega_+,\;\omega_-)$ and $x(\omega_+,\;\omega_-)$:
\begin{equation}
    \left(\begin{array}{c}dt\\ dx\end{array}\right)=\left(\begin{array}{cc}
        \frac{\partial t}{\partial \omega_+} & \frac{\partial t}{\partial \omega_-} \\
       \frac{\partial x}{\partial \omega_+}  & \frac{\partial x}{\partial \omega_-}
    \end{array}\right)\; \left(\begin{array}{c}d\omega_+\\ d\omega_-\end{array}\right)\;.
    \label{eqn totaldiff}
\end{equation}
As long as the Jacobian 
\begin{equation}
    J=\frac{\partial t}{\partial \omega_+}\;\frac{\partial x}{\partial \omega_-}-\frac{\partial t}{\partial \omega_-}\;\frac{\partial x}{\partial \omega_+}\label{eqn jacobianCartesianToCharac}
\end{equation} 
does not vanish --- which is true in regions of hyperbolicity and without caustics --- it is possible to change independent variables $(\omega_+,\;\omega_-)\rightarrow (t,\;x)$. 
Solving Eq.~(\ref{eqn totaldiff}) in terms of 
$d\omega_\pm$, we obtain the relations
\begin{equation}
       \frac{\partial \omega_+ }{\partial t}=\frac{1}{J}\frac{\partial x}{\partial \omega_-}  ~~~~~~~~~~~~~  \frac{\partial \omega_+ }{\partial x}=-\frac{1}{J}\frac{\partial t}{\partial \omega_-} \;,\label{eqn cartesianToCharacCoords1}
\end{equation}

\begin{equation}
       \frac{\partial \omega_- }{\partial t}=-\frac{1}{J}\frac{\partial x}{\partial \omega_+}  ~~~~~~~~~~~~~  \frac{\partial \omega_- }{\partial x}=\frac{1}{J}\frac{\partial t}{\partial \omega_+} \;.\label{eqn cartesianToCharacCoords2}
\end{equation}
Using these in Eq.~\eqref{char}, one obtains the alternative definition of characteristics as surfaces of constant $\omega_\pm$ in the $(t,x)$-plane, i.e., $\omega_{\pm}(t,x)=\text{constant}$, defined by the equations
\begin{equation}
       \frac{\partial \omega_+ }{\partial t}+\xi_-\;\frac{\partial \omega_+}{\partial x}=0  ~~~~~~~~~~~~~  \frac{\partial \omega_- }{\partial t}+\xi_+\;\frac{\partial \omega_-}{\partial x}=0 \;.\label{slopes}
\end{equation}
Note, however, that the coordinate transformation $(\omega_+,\;\omega_-)\rightarrow (t,\;x)$ may become ill-defined, which may correspond to caustic formation or loss of hyperbolicity, as discussed in Secs.~\ref{no} and~\ref{cuspy}.

For a given $P(X)$ one can integrate Eq.~\eqref{ode} numerically or analytically and obtain the solutions for $\tau$ and 
$\chi$ in terms of $\omega_{+}$ and $\omega_{-}$. It is convenient to introduce some useful notations beforehand. The characteristic slopes (phase velocities) $\xi_{\pm}$ can be rewritten in the following elegant form~\cite{Babichev:2016hys}:
\begin{equation}
\label{csv}
\xi_{+} = \frac{v+c_s}{1 + vc_s} \qquad \xi_{-} = \frac{v-c_s}{1 - vc_s}\; ,
\end{equation}
where $c_s$ is the sound speed~\cite{Garriga:1999vw}:
\begin{equation} 
\label{cs}
c_s =\sqrt{\frac{P_X}{P_X+ 2X P_{XX} }} \; ,
\end{equation}
and 
\begin{equation}
\label{v}
v \equiv -\frac{\chi}{\tau} \; .
\end{equation}
Note that Eq.~\eqref{csv} has the form of 
the relativistic velocity addition law (hence, the notation $v$). The integrals of Eq.~\eqref{ode} can be written as
\begin{equation}
\label{integrated}
\int \frac{dX}{c_s X} - \ln \left(\frac{1+v}{1-v} \right)=C_{-} (\omega_{+}) \qquad \int \frac{dX}{c_s X} + \ln \left(\frac{1+v}{1-v} \right)=C_{+} (\omega_{-}) 
 \; ,
\end{equation}
where $C_{+}$ and $C_{-}$ are the so called Riemann invariants. 
For given $C_{+}$ and $C_{-}$, one can express $\tau$ and $\chi$ as functions of $\omega_{-}$ and $\omega_{+}$, and then produce characteristic curves using Eqs.~\eqref{char} and~\eqref{csv}. For $C_{+}$ and $C_{-}$ being both non-constant, i.e., for {\it generic waves}, the phase velocities $\xi_{+}$ and $\xi_{-}$ depend on both coordinates $\omega_{+}$ and $\omega_{-}$, i.e., $\xi_{+}=\xi_{+} (\omega_{+}, \omega_{-})$ and $\xi_{-}=\xi_{-} (\omega_{+}, \omega_{-})$. (The trivial scenario $P(X)=X$ and DBI are notable exceptions, as we will see below.) In the particular 
case, when one of Riemann invariants, $C_{+}$ or $C_{-}$, is chosen to be constant, we deal with the so called {\it simple waves}. In this case, the solution in the $(\tau, \chi)$-plane collapses to a line segment; this greatly simplifies the proof 
of some statements, e.g., caustic formation in generic $P(X)$-theories~\cite{Babichev:2016hys}, as discussed in the end of Sec.~\ref{no}. Unless the opposite is stated, 
we will deal with generic waves in what follows.

Let us apply the method of characteristics to the case $P(X)=X$. Despite being trivial, this scenario is interesting, because it shares certain similarities with the model of our major interest --- DBI. Upon substituting $c_s=1$ into Eq.~\eqref{integrated}, we get 
\begin{equation}
\tau-\chi =\mbox{exp} \left(\frac{C_+ (\omega_{-})}{2} \right) \qquad \tau +\chi=\mbox{exp} \left(\frac{C_{-} (\omega_{+})}{2} \right) \; .
\end{equation}
This gives $\xi_{\pm} =\pm 1$ by virtue of Eq.~\eqref{ode}. Hence, one can write
\begin{equation}
\label{trivialx}
x (\omega_{+}, \omega_{-})=X_{+} (\omega_{+})+X_{-} (\omega_{-}) \; ,
\end{equation}
and 
\begin{equation}
\label{trivialt}
t(\omega_{+}, \omega_{-})=X_{+} (\omega_{+})-X_{-} (\omega_{-}) \; ,
\end{equation}
where $X_{+} (\omega_{+})$ and $X_{-} (\omega_{-})$ are arbitrary functions. It is natural to set $X_{\pm}= \omega_{\pm}/2$, which corresponds to the choice of $\omega_{\pm}$ as lightcone coordinates, $\omega_{\pm}=x \pm t$. Consequently, one gets 
\begin{equation}
\label{trivial}
\tau =f_1 (t+x)+f_2 (t-x) \qquad \chi=f_1 (t+x)-f_2 (t-x) \; ,
\end{equation}
where $f_1 \equiv 1/2 \cdot \mbox{exp} \left(C_{-} /2 \right)$ and $f_2 \equiv 1/2 \cdot \mbox{exp} \left(C_+/2 \right)$.
Of course, the result~\eqref{trivial} describing generic waves in the canonical case $P(X)=X$ could be anticipated from the beginning.

\section{DBI in 2D flat spacetime}
\label{dbicharacteristics}  

We apply the machinery described in the previous section to the case of DBI. The DBI sound speed inferred from Eq.~\eqref{cs} upon substituting $P(X)=-\sigma \sqrt{1-2X}$ reads 
\begin{equation}
\label{csdbi}
c_s =\sqrt{1-2X} \; . 
\end{equation}
Substituting the latter into Eq.~\eqref{integrated} and using Eq.~\eqref{csv}, one gets
\begin{equation}
\label{eqmajordbi}
  \ln \left(\frac{1 - \xi_{+}}{1 + \xi_{ +}} \right) =C_{-} (\omega_{+}) \qquad  \ln \left(\frac{1 + \xi_{-}}{1 - \xi_{ -}} \right) =C_{+} (\omega_{-}) \; .
\end{equation}
Hence, one can express the characteristic slopes as 
\begin{equation}
\xi_{+}=- \tanh \left(\frac{C_{-} (\omega_{+})}{2} \right) \qquad \xi_{-}= \tanh \left(\frac{C_{+} (\omega_{-})}{2} \right)\; .\label{reldbi}
\end{equation}
Note that the Riemann invariants in DBI are subject to the constraint 
\begin{equation}
C_{-} (\omega_{+})+C_{+} (\omega_{-})<0 \; ,
\end{equation}
which is a direct consequence of the inequality $\xi_{+}>\xi_{-}$. This is in line with the intuition that $\xi_{+}$- and $\xi_{-}$-characteristics correspond to right- and left-moving particles, respectively. More strictly, the inequality $\xi_{+}>\xi_{-}$ can be obtained from Eq.~\eqref{csv}, which yields
\begin{equation}
\label{cusper}
\xi_{+}-\xi_{-}=\frac{2c_s}{1+\chi^2}  \; ,
\end{equation}
and from $c_s>0$.  

Crucially, we observe that
\begin{equation}
\label{ce}
\xi_{+} =\xi_{+} (\omega_{+}), \qquad \xi_{-}=\xi_{-} (\omega_{-}) \; .
\end{equation}
This is in contrast to generic $P(X)$-theories, where each $\xi$ generically depends on both $\omega$'s.
The systems fulfilling the property~\eqref{ce} are called totally linearly degenerate systems~\cite{Majda} 
and they have remarkable properties regarding wave propagation, see Sec.~\ref{no}. Among $P(X)$-theories with a finite sound speed, only DBI and the trivial case $P(X)=X$ 
fulfill this property. This follows from 
\begin{equation}
\label{interesting}
\frac{\partial \xi_{\pm}}{\partial \omega_{\mp}} =\frac{4 \, c^4_s \, X^3}{(1 \pm vc_s)^3 \, \tau^3 \,  P^2_{X}} \cdot \left[P_{XXX} P_X -3P^2_{XX}  \right] \cdot \frac{\partial \chi}{\partial \omega_{\mp}} \; ,
\end{equation}
which can be obtained by manipulating Eqs.~\eqref{ode},~\eqref{csv}, and~\eqref{cs}. It is straightforward to show that the quantity in the square brackets on the r.h.s. vanishes identically only for $P(X)=X$ and $P(X)=-\sqrt{c_1 +c_2 X}$ (DBI), where $c_1 \neq 0$ and $c_2$ are constants\footnote{The case $c_1=0$ dubbed cuscuton~\cite{Afshordi:2006ad} is special, as it corresponds to an infinite sound speed~\eqref{cs}.}, cf. Refs.~\cite{Deser:1998wv, Boillat:1970gw}.

\begin{figure}[!htb]
\begin{center}
    \includegraphics[width=\textwidth]{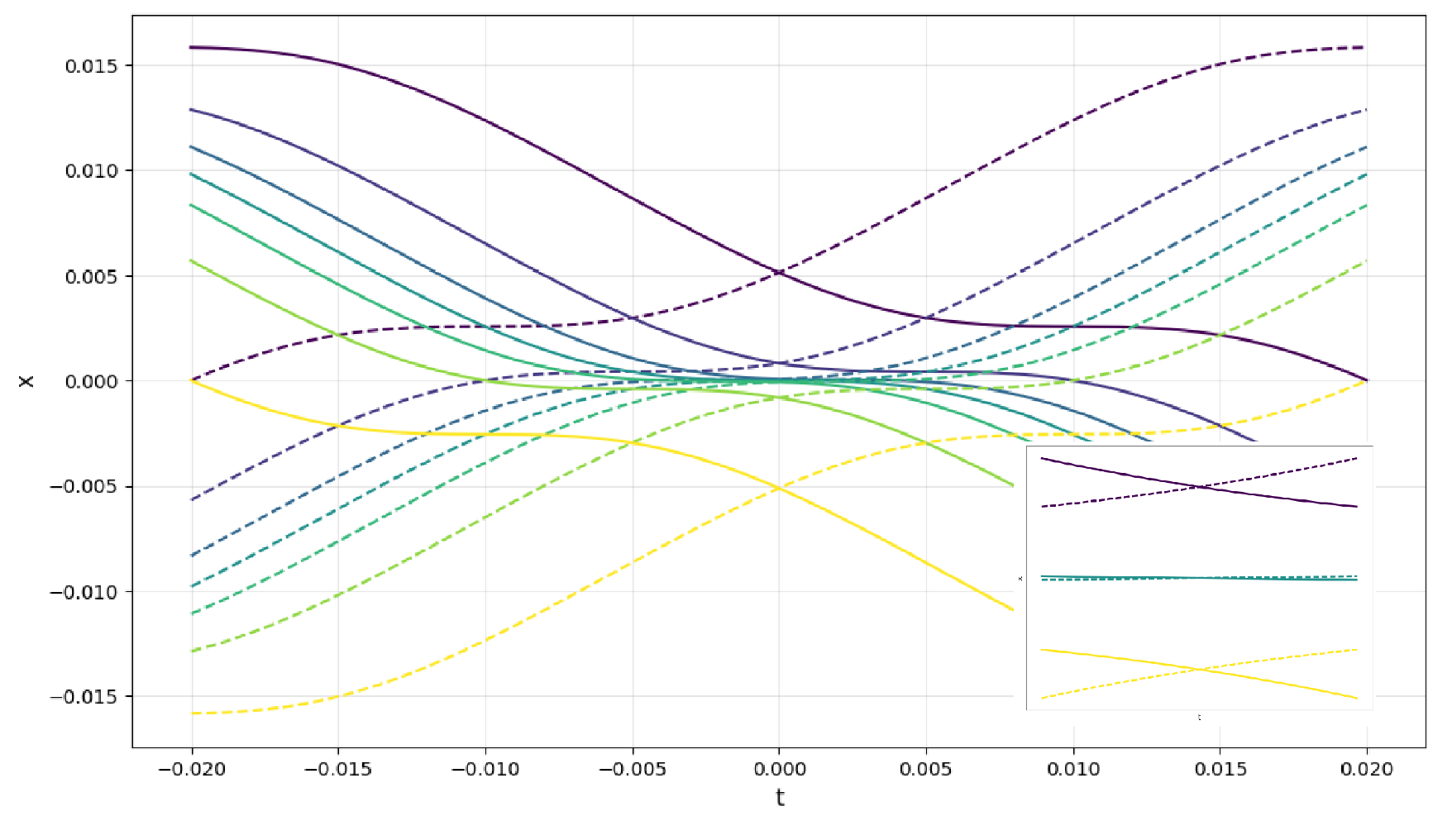} 
\end{center}
    \caption{Characteristic curves in DBI in 2D flat spacetime are demonstrated for a particular generic wave. The $\xi_{-}$- and $\xi_{+}$-characteristics are depicted with a solid and dashed lines, respectively. The Riemann invariants have been set to $C_{\mp} (\omega_{\pm})=-0.003 -\sin^2 (100 \omega_{\pm})$. With this choice primarily made for the clarity of presentation, one also warrants that hyperbolicity is fulfilled. The time $t$ is expressed in terms of 
    $\omega_{+}$ and $\omega_{-}$ as $t=\omega_{+}-\omega_{-}$, which corresponds to regular initial conditions for the DBI scalar field $\phi$. One observes that characteristics belonging to the same family do not cross, i.e., caustics do not form. In particular, 
    this holds in the central region near $x =0$ and $t=0$, where characteristics come close to each other, as it is clear from the zoomed picture 
    in the right bottom corner. 
    } \label{no_caustics}
\end{figure}

By virtue of Eq.~\eqref{ce}, in the hyperbolic case one can write down the solution of Eq.~\eqref{char} as follows (see the derivation below): 
\begin{equation}
\label{sum1}
x(\omega_{+}, \omega_{-})=X_{+} (\omega_{+}) +X_{-} (\omega_{-}) \; ,
\end{equation}
and 
\begin{equation}
\label{sum2}
t (\omega_{+}, \omega_{-})=T_{+} (\omega_{+})- T_{-} (\omega_{-}) \; ,
\end{equation}
where the functions $X_{\pm}$ and $T_{\pm}$ are related to $\xi_{\pm}$ by $(dX/dT)_{\pm}=\pm \xi_{\pm}$. Let us point out here the remarkable similarity between DBI and the truly linear case $P(X)=X$ discussed in the end 
of the previous section. Recalling the relation~\eqref{reldbi}, we write
\begin{equation}
\label{tbi}
\left( \frac{d X}{d T} \right)_{+} =\xi_{+}=- \tanh \left( \frac{C_{-} (\omega_{+})}{2} \right) \qquad \left( \frac{d X}{d T} \right)_{-} =-\xi_{-}=- \tanh \left( \frac{C_{+} (\omega_{-})}{2} \right) \; . 
\end{equation}
By choosing arbitrary $C_{\pm} (\omega_{\mp})$ and $T_{\pm} (\omega_{\pm})$, one can get $X_{\pm}$ upon integrating Eq.~\eqref{tbi}. In Fig.~\ref{no_caustics} we demonstrate characteristics generated for the particular choice of $C_{\pm}$ and $T_{\pm}$.

Let us prove Eqs.~\eqref{sum1} and~\eqref{sum2}. For this purpose we notice that $x(\omega_+,\omega_-)$ can be written in the form
  \begin{equation}
x(\omega_+,\omega_-) =X_-(\omega_-)+\int d\omega_+\,\xi_{+} (\omega_{+})\;\frac{\partial t(\omega_+,\omega_-)}{\partial \omega_+}\;, \label{eqn solx1}
 \end{equation}
 which is obtained from $(dx/dt)_{+}=\xi_{+} (\omega_{+})$. Similarly, from $(dx/dt)_-=\xi_{-} (\omega_{-})$ we obtain
  \begin{equation}
x(\omega_+,\omega_-) =X_+(\omega_+)+\int d\omega_-\,\xi_{-} (\omega_{-})\;\frac{\partial t(\omega_+,\omega_-)}{\partial \omega_-}\;. \label{eqn solx2}
 \end{equation}
The consistency condition 
\begin{equation}
\frac{\partial}{\partial \omega_{+}} \frac{\partial x}{\partial \omega_{-}}=\frac{\partial}{\partial \omega_{-}} \frac{\partial x}{\partial \omega_{+}} 
\end{equation}
then tells us that
 \begin{equation}
 \label{separability}
     (\xi_+ -\xi_-)\;\frac{\partial^2 t(\omega_+,\omega_-)}{\partial \omega_+\,\partial \omega_-}=0\;.
 \end{equation}
As we are interested in regions where the equation of motion is hyperbolic (the non-hyperbolic case is to be considered in Sec.~\ref{cuspy}), two families of characteristics always have different phase velocities $\xi_+\neq \xi_-$, thus we find that
\begin{equation}
\frac{\partial^2 t}{\partial \omega_+\,\partial \omega_-}=0 \; , 
\end{equation}
which proves Eq.~\eqref{sum2}. From Eqs.~(\ref{eqn solx1}) and (\ref{eqn solx2}) we see that 
\begin{equation}
x (\omega_{+}, \omega_{-})= \int d\omega_+\,\xi_{+} (\omega_{+})\;\frac{d T_+(\omega_+)}{d \omega_+}-\int d\omega_-\,\xi_{-} (\omega_{-})\;\frac{d T_-(\omega_-)}{d \omega_-} \; ,
\end{equation}
which justifies Eq.~\eqref{sum1}. We stress that this simplification is possible in DBI in 2D, because of the property discussed above, i.e., $\xi_{\pm}=\xi_{\pm}(\omega_\pm)$.

Now let us write the solution for $\tau$ and $\chi$ for given $C_{+}$ and $C_{-}$. Notably, it can be expressed in the analytical form. By virtue of Eqs.~\eqref{ode} and~\eqref{ce}, one finds that the solution fulfills
$\tau=-\xi_- (\omega_{-}) \chi+b(\omega_{-})$, where $b(\omega_{-})$ is some function of $\omega_{-}$. The latter is fixed from the consistency with Eq.~\eqref{csv} to be 
$b(\omega_{-})=\sqrt{1-\xi^2 (\omega_{-})}$. The second of Eq.~\eqref{ode} yields the same modulo the obvious replacement $\xi_{-} (\omega_{-}) \rightarrow \xi_{+} (\omega_{+})$. Consequently, the solution must satisfy
 \begin{equation}
 \label{solutions}
 \tau=-\xi_{-} (\omega_{-}) \chi+ \sqrt{1-\xi^2_{-} (\omega_{-})} =-\xi_{+} (\omega_{+}) \chi +\sqrt{1-\xi^2_{+} (\omega_{+})} \; .
 \end{equation}
 In combination with Eq.~\eqref{reldbi}, these can be used to express $\tau$ and $\chi$ in terms of $\omega_{+}$ and $\omega_{-}$:
 \begin{equation}
 \tau=\frac{\sinh \frac{C_- (\omega_{+})}{2} +\sinh \frac{C_+ (\omega_{-})}{2}}{\sinh \left(\frac{C_{+} (\omega_{-})+C_{-} (\omega_{+}}{2}) \right)}, \qquad 
\chi=\frac{\cosh \left(\frac{C_- (\omega_{+})}{2}  \right) -\cosh \left(\frac{C_+ (\omega_{-})}{2} \right) }{\sinh \left(\frac{C_{+} (\omega_{-}) +C_{-} (\omega_{+})}{2} \right)} \; .
 \end{equation}
We observe that DBI is an exactly solvable model in 2D flat spacetime. 
This is a manifestation of Nambu-Goto integrability in this setting, which is also 
clear from the study of this model in the conformal gauge~\cite{Blanco-Pillado:2025gzs}, see Appendix A.

\section{No caustics in DBI in 2D flat spacetime\\ (hyperbolic case)}
\label{no}

It is known that hyperbolic systems fulfilling the property~\eqref{ce} are caustic free, see, e.g., Refs.~\cite{lax1964development, lax1973hyperbolic, lax1955xii}. We present the proof of this statement below in this section. This proof is rather straightforward, and it is based on Eq.~\eqref{sum2}, which is a direct consequence of Eq.~\eqref{ce}. We also briefly discuss a more generic approach to the study of caustics by Lax~\cite{lax1964development, lax1973hyperbolic, lax1955xii} in Appendix B.  

Before giving the proof, let us describe the relation of caustics to the properties of the coordinate transformation $\omega_{\pm}=\omega_{\pm} (t,x)$. Recall that $\omega_{+}$ and $\omega_{-}$ are natural coordinates suggested by the characteristic method used to solve PDE~\eqref{partial}. This transformation is described by the Jacobian
 \begin{equation}
 \label{jacob}
 \tilde{J}=\frac{\partial \omega_{+}}{\partial t} \frac{\partial \omega_{-}}{\partial x}-\frac{\partial \omega_{-}}{\partial t} 
 \frac{\partial \omega_{+}}{\partial x} =\left(\xi_{+}-\xi_{-} \right)  \cdot \frac{\partial \omega_{+}}{\partial x} \frac{\partial \omega_{-}}{\partial x} \; ,
 \end{equation}
 where we have used Eq.~\eqref{slopes} in the second equality. Generally characteristics may intersect meaning that one point in the $(t,x)$-plane corresponds to multiple $(\omega_{+}, \omega_{-})$. At this point the Jacobian $J$ in Eq.~\eqref{eqn jacobianCartesianToCharac} vanishes, or equivalently $\tilde{J}$ diverges, $\tilde{J} \rightarrow \infty$ \cite{Orlov}. This singularity taking place at characteristics crossing reflects caustic formation. Caustics are not a mere mathematical artefact pertaining to a particular choice of coordinates. As it has been noted in Sec.~\ref{method}, characteristics have a clear physical meaning: they correspond to the trajectories of particles in the high momentum limit. At their intersection the field derivatives $\tau$ and $\chi$ become multi-valued, and consequently the second derivatives of the scalar field $\phi$ blow up generically, as it follows, e.g., from (see also a relevant discussion in~Ref.~\cite{Babichev:2020tct})
 \begin{equation}
 \label{dtau}
 \frac{\partial \chi}{\partial x}=\frac{\partial \chi}{\partial \omega_{+}} \frac{\partial \omega_{+}}{\partial x}+\frac{\partial \chi}{\partial \omega_{-}} \frac{\partial \omega_{-}}{\partial x} \; .
 \end{equation}
 Note also that the coordinate transformation becomes non-invertible at $\xi_{+}=\xi_{-}$ corresponding to loss of hyperbolicity. Caustics (referred to as cusps) appear in this case as well, but we postpone their discussion till Sec.~\ref{cuspy}.

Here we proceed assuming that $\xi_{+} \neq \xi_{-}$. It is convenient to express $\partial \omega_{\pm}/\partial x$ in terms of $\partial t/\partial \omega_{\pm}$. Using Eqs.~\eqref{eqn cartesianToCharacCoords1} and~\eqref{eqn cartesianToCharacCoords2}, we obtain 
\begin{equation}
\label{wt}
\frac{\partial \omega_{\pm}}{\partial x} =\frac{1}{\xi_{\pm}-\xi_{\mp}} \cdot \frac{1}{\partial t/\partial \omega_{\pm}} \; .
\end{equation}
As it follows, the condition $\partial \omega_{\pm}/\partial x \rightarrow \infty$ corresponding to caustic formation can be reinterpreted as $\partial t/\partial \omega_{\pm} \rightarrow 0$. Substituting~ Eq.~\eqref{wt} into Eq.~\eqref{dtau}, one gets
\begin{equation}
\label{ddtaut}
\frac{\partial \chi}{\partial x}=\frac{1}{\xi_{+}-\xi_{-}} \cdot \left[\frac{\partial \chi}{\partial \omega_{+}} \cdot \frac{1}{\partial t/\partial \omega_{+}}-\frac{\partial \chi}{\partial \omega_{-}} \cdot \frac{1}{\partial t/\partial \omega_{-}} \right] \; .
\end{equation}
The similar expressions 
hold for $\partial \tau/\partial t$ and $\partial \chi/\partial t$. One observes that the second derivatives of the field $\phi$ generically become infinite at caustics, i.e., in the limit $\partial t/\partial \omega_{\pm} \rightarrow 0$.

{\it Proof of no caustics.} Let us argue that the caustic singularity manifested at $\partial t/\partial \omega_{\pm}=0$ cannot appear in DBI, if one assumes smooth initial 
conditions. The no-caustic proof relies on the relation $\partial^2 t/\partial \omega_+ \partial \omega_{-}=0$, a property of DBI established in the previous section, which follows from $\xi_{\pm}=\xi_{\pm} (\omega_{\pm})$. This implies in particular that $\partial t/\partial \omega_{+}$ remains constant along a $\xi_{-}$-curve. Hence, if $\partial t/\partial \omega_{+}=0$ at some point, this value is conserved along the $\xi_{-}$-characteristic, which contains this point and crosses the surface $t=t_0$ where initial conditions are set. Having $\partial t/\partial \omega_{+}=0$ at $t=t_0$ implies infinite second order derivatives initially by virtue of Eq.~(\ref{wt}). The same argument is applicable to the quantity $\partial t/\partial \omega_{-}$. In other words, non-zero initial values of $\partial t/\partial \omega_{\pm}$ remain so at later times. This proves that caustics do not form starting from a regular scalar field profile in DBI. Notably, this happens despite that phase velocities $\xi_{\pm}$ are not constants\footnote{In this regard, we disagree with the approach 
of Ref.~\cite{Pasmatsiou:2017vcw}, which uses constancy of $\xi$ as a criteria for caustic formation in extended DBI.}, as is clearly illustrated in Fig.~\ref{no_caustics}. Unlike in Ref.~\cite{Mukohyama:2016ipl} focusing on the particular case of simple waves, our proof holds for generic waves. 
We would like to reiterate that the discussion so far has been limited to DBI in 2D flat spacetime. In the next section 
we demonstrate how DBI resists to caustics even beyond this simplified scenario.

It is instructive to compare the case of DBI with generic $P(X)$-theories in 2D flat spacetime, and demonstrate how the latter naturally lead to caustic formation. Generically, we should write
\begin{equation}
    \frac{\partial^2 t}{\partial \omega_{+} \partial \omega_{-}}{}=\frac{1}{(\xi_+-\xi_-)}\Big(\frac{\partial \xi_-}{\partial \omega_+}\frac{\partial t}{\partial \omega_-}- 
    \frac{\partial \xi_+}{\partial \omega_-}\frac{\partial t}{\partial \omega_+}\Big)\,.\label{eqn cond2}
\end{equation}
To obtain this, one follows the same steps, which have led to Eq.~\eqref{separability}, but considers $\partial \xi_{\pm}/\partial \omega_{\mp} \neq 0$, see Eq.~\eqref{interesting}. When $\partial t^2/\partial \omega_{+} \partial \omega_{-} \neq 0$, it may be possible to get $\partial t/\partial \omega_{\pm} =0$ corresponding to caustic formation starting from smooth initial conditions. To show that this is indeed the case in generic $P(X)$-theories, we consider simple waves, which fulfill the property that $\tau$ and $\chi$ (and hence $\xi_{+}$ and $\xi_{-}$) depend only on $\omega_{+}$ {\it or} $\omega_{-}$. For concreteness, we assume the dependence on $\omega_{+}$, so that $\partial \xi_{-}/\partial \omega_{+} \neq 0$, while $\partial \xi_{+}/\partial \omega_{-}=0$. Recall that such a solution is achieved by setting $C_{+} (\omega_{-})=\mbox{const}$ in Eq.~\eqref{integrated}. Integrating Eq.~\eqref{eqn cond2} along a $\xi_{-}$-characteristic curve, so that $\omega_{+} =\mbox{const}$, we obtain
 \begin{equation}
 \label{gencaus}
 \frac{\partial t}{\partial \omega_{+}} \Bigr|_{t}=\frac{\partial t}{\partial \omega_{+}} \Bigr|_{t_0} +\frac{\partial \xi_{-}}{\partial \omega_{+}}\int^{t}_{t_0} \frac{dt'}{\xi_{+}-\xi_{-}} \; .
 \end{equation}
 Here we have taken into account that $\partial \xi_{-}/\partial \omega_{+}$, --- a function of $\omega_{+}$ only by the assumption 
 of a simple wave, --- remains constant along the $\xi_{-}$-curve. Recall that we assume hyperbolicity to be strictly fulfilled here, i.e., $\xi_{+} \neq \xi_{-}$. One infers from Eq.~\eqref{gencaus} 
 that $\partial t/\partial \omega_{+}$ indeed can turn into zero after a finite time $t$ meaning crossing of $\xi_{-}$-characteristics, provided that two terms on the r.h.s. have different signs. This could be understood in a more straightforward way from the fact that 
 $\xi_{-}$-characteristics are straight non-parallel lines for the simple wave~\cite{Babichev:2016hys}. However, the analysis based on Eq.~\eqref{eqn cond2} proves to be useful also in the situation, when no simple wave solution exists. We will deal with such a situation in the next section. There we will observe that violation of $\partial \xi_{\pm}/\partial \omega_{\mp} = 0$ does not
 warrant formation of caustics.

\section{Towards realistic domain walls:\\ beyond DBI in 2D flat spacetime}
\label{beyond}

As discussed in the previous sections, a defining feature of DBI in 2D flat spacetime is the dependence of the slopes $\xi_{\pm}$ only on the respective parameter along the characteristic curve, i.e., $\xi_+=\xi_+(\omega_+)$ and $\xi_-=\xi_-(\omega_-)$. Here we demonstrate that this property is violated beyond the basic scenario. We are primarily interested in three physically relevant cases, i.e., the case of $D>2$ spacetime, the case of the expanding Universe, 
and the case of linearly extended DBI. Remarkably, however, the caustic free nature of DBI in the hyperbolic case persists through these and other modifications.

The relevant equation of motion generalizing three cases mentioned above, can be written in the following form:
\begin{equation}
\label{dbi12}
A \dot{\tau}+2B \tau'+C \chi'=Q \; .
\end{equation}
Here the coefficients $A, B$, and $C$ take the same form as in the original DBI in Minkowski 2D space, i.e., they are given by Eq.~\eqref{ABC} with $P(X)=-\sigma \sqrt{1-2X}$. The difference from the 
scenario considered in the past sections is in the term $Q \neq 0$ arising on the r.h.s. Note that conclusions of this section 
hold for an arbitrary smooth function $Q =Q (t, x, \tau, \chi)$. 
Crucially, we assume that $Q$ does not depend on the derivatives of $\tau$ and $\chi$. This indeed holds in all three cases 
mentioned above, and we comment more on it later in this section.
Repeating the steps of Sec. 3, one applies the characteristic method for solving Eq.~\eqref{dbi12}, and results with two families of curves defined by the phase velocities $\xi_{+}$ and $\xi_{-}$. These are again given by the quadratic equation $A\xi^2 -2B \xi +C=0$. Hence, one can use the same expressions~\eqref{csv} for $\xi_{+}$ and $\xi_{-}$ in terms of $\tau$ and $\chi$ as in DBI in 2D flat spacetime. However, the dependence of $\xi_{+}$ and $\xi_{-}$ on $\omega_{+}$ and $\omega_{-}$ gets modified, as will become clear shortly. 

The first two characteristic equations~\eqref{char} remain intact, i.e., $( \partial x/\partial \omega)_{\pm} =\xi_{\pm} (\partial t /\partial \omega)_{\pm}$. On the other hand, characteristic equations for the fields $\tau$ and $\chi$ take the form 
\begin{equation}
\label{dbimod}
 \frac{\partial \tau}{\partial \omega_{\pm}}  +\xi_{\mp} \frac{\partial \chi}{\partial \omega_{\pm}} =\frac{Q}{A\xi_{\pm}} \cdot \frac{\partial  x}{\partial \omega_{\pm}}  \; .
\end{equation}
With the use of Eqs.~\eqref{csv},~\eqref{v}, and~\eqref{csdbi}, this can be rewritten as
\begin{equation}
\label{dbi12ksi}
\frac{\partial}{\partial \omega_{\mp}} \ln \left( \frac{1 \mp \xi_{\pm}}{1 \pm \xi_{\pm}} \right)=\frac{2 Q (1\pm c_s v)}{c_s \tau  \,  (1-v^2) A\xi_{\mp}} \cdot \left(\frac{\partial x}{\partial \omega_{\mp}} \right) \; .
\end{equation}
It is convenient to recast the latter into a somewhat more tractable form using Eqs.~\eqref{ABC},~\eqref{char},~\eqref{csv}, and~\eqref{v}:
\begin{equation}
\label{simpl3dm}
\frac{\partial \xi_{\pm}}{\partial \omega_{\mp}}=\mp \frac{Q \tau c^2_s \cdot (1-v^2)}{\sigma (1 \pm c_s v) (1+\chi^2)} \cdot \left(\frac{\partial t}{\partial \omega_{\mp}} \right) \; .
\end{equation}
This means that $\partial \xi_{\pm}/\partial \omega_{\mp} \neq 0$ generically, hence characteristics cease being parallel upon including $Q \neq 0$.

Nevertheless, one can prove that DBI remains protected against caustics (in the hyperbolic case). Crucial for this proof is the observation from Eq.~\eqref{simpl3dm} that $\partial \xi_{\pm}/\partial \omega_{\mp} \rightarrow 0$ exactly where 
caustics are expected to form, i.e., in the limit $\partial t/\partial \omega_{\pm} \rightarrow 0$. This precludes crossing of characteristics, as we will see shortly. For this purpose, one considers $\partial^2 t/\partial \omega_{+} \partial \omega_{-}$ given by the generic expression~\eqref{eqn cond2}. Recall that this quantity vanishes in DBI in 2D flat spacetime, while 
for $Q \neq 0$ we obtain upon substituting Eq.~\eqref{simpl3dm} into Eq.~\eqref{eqn cond2}: 
\begin{equation}
\label{ddtQ}
\frac{\partial^2 t}{\partial \omega_{+} \partial \omega_{-}}=\frac{Q c_s \tau}{\sigma (1+\chi^2)} \cdot \frac{\partial t}{\partial \omega_{+}} \frac{\partial t}{\partial \omega_{-}} \; .
\end{equation}
We pick an arbitrary $\xi_{-}$-characteristic for concreteness, and consider evolution of the quantity $\partial t/\partial \omega_{+}$ along this 
curve. The above equation can be then considered as a \textit{linear} ordinary differential equation of first order on $\partial t/\partial \omega_{+}$ as a function of $\omega_-$. 
Now, if one assumes that there is a caustic, $\partial t/\partial \omega_{+}=0$, at one point, then $\partial t/\partial \omega_{+}=0$ everywhere, by uniqueness of the solution for linear ODE. But by assumption the initial conditions do not allow a caustic, therefore we arrived at the contradiction. This argument shows that no caustics develop for modifications of the original DBI equation of motion of the form~(\ref{dbi12}), in regions of hyperbolicity. It should be stressed that in the above derivation we implicitly assumed regularity of first derivatives of the field $\phi$ and of $\partial t/\partial \omega_\pm$ (otherwise one would deal with infinite rarefaction).

{\it DBI in 3D.} The characteristic method reviewed in Sec.~\ref{method} is suitable for solving PDEs in 2D. One can analyze the problem in $D>2$, if there is a spherical symmetry, so that one effectively lives in 2D, and the radius 
$r$ is the only relevant spatial coordinate, i.e., $x \equiv r$. However, the equation describing evolution of spherical waves in DBI is modified compared to the case of planar waves in 2D. It takes the form~\eqref{dbi12}, where the function $Q$ is given by
\begin{equation}
Q=\frac{(D-2)\sigma \chi}{c_s r} \; .
\end{equation}
As it has been discussed above, the pattern of characteristic curves changes in the presence of $Q \neq 0$. 
This is at odds with the discussion of Ref.~\cite{deRham:2016ged} neglecting the $1/r$ term in the equation of motion for the scalar $\phi$. 
 The confusion may come from the aforementioned fact that the expressions for $\xi_{\pm}$ in terms of $\tau$ and $\chi$ remain unaffected by the presence of the $1/r$-term. Still this term cannot be ignored, because it changes the dependence of $\tau$ and $\chi$, and consequently $\xi_{\pm}$, on $\omega_{\pm}$, as has been proven earlier.
 On the other hand, we agree with Ref.~\cite{deRham:2016ged} on the absence of DBI caustics in spherical waves in the hyperbolic case. Yet, the presence of $Q \neq 0$ may affect formation of non-hyperbolic caustics (cusps) discussed in the next section.

{\it DBI in the expanding Universe.} As we are primarily interested in cosmic domain walls, it is natural to include the 
Universe expansion in the analysis. 
In that case, one starts with the action~\eqref{dbireal}. We again restrict to the case of DBI in 2D, i.e., neglect dependence of the 
DBI field on the coordinate $y$, or equivalently limit to planar waves on a wall. In this setup, the field $\phi$ satisfies Eq.~\eqref{dbi12} with $Q$ given by
\begin{equation}
Q=-\frac{3\sigma {\cal H}}{c_s} \tau \; ,
\end{equation}
where ${\cal H}$ is the conformal Hubble rate, ${\cal H} \equiv \partial \ln a/\partial \eta$, and $\eta$ is the conformal time. Note that here and in Eq.~\eqref{dbi12} the time derivative should be taken with respect to $\eta$, i.e., $\tau \equiv \partial \phi/ \partial \eta$, and $\dot{\tau} \equiv \partial \tau/\partial \eta$. Obviously, the case of the expanding Universe is not exceptional in a sense that one can consider any curved background, and the conclusion about the absence of caustics will still hold (assuming hyperbolicity).

{\it DBI extended by a linear term.} Finally, we note that the so called domain wall problem~\cite{Zeldovich:1974uw} motivates a slight modification of DBI, which also leads to $Q \neq 0$. 
Namely, in the expanding Universe the energy density of domain walls 
redshifts too slowly relative to the background matter, and they tend to dominate cosmic expansion. 
The problem is solved, if domain walls are unstable and annihilate at some point. 
This can be achieved by an explicit breaking of $Z_2$-symmetry, i.e., one considers the following modification of the action~\eqref{dw}: 
\begin{align}
    S_{DW}=\int d^4 x \sqrt{-g} \left[ -\frac{1}{2}g^{\mu \nu} \partial_{\mu} \Psi \partial_{\nu} \Psi-\frac{1}{4}\lambda\left(\Psi^2-\langle \Psi \rangle^2\right)^2+ \frac{\epsilon \Psi^m}{2\langle \Psi \rangle^m} \right] \, , \label{eqn source pi}
\end{align}
where the exponent $m$ is {\it odd}, and $\epsilon$ is the coefficient assumed to be constant. In the thin wall limit in the expanding Universe the action~\eqref{eqn source pi} reduces to 
\begin{equation}
\label{dbilinear}
S_{lin-DBI}=- \int d \eta d x d y a^3  \left[ \sigma \sqrt{1-2X}-\epsilon a \phi \right] \; .
\end{equation}
We discuss details of derivation of Eq.~\eqref{dbilinear} from Eq.~\eqref{eqn source pi} in Appendix C. The equation of motion for the field $\phi$ is given by Eq.~\eqref{dbi12}, where 
\begin{equation}
Q=\epsilon \; ,
\end{equation}
and we again simplified to the case of DBI in 2D flat spacetime.

\section{Caustic formation in the non-hyperbolic case}
\label{cuspy}

\begin{figure}[!htb]
\begin{center}
    \includegraphics[width=0.3\textwidth]{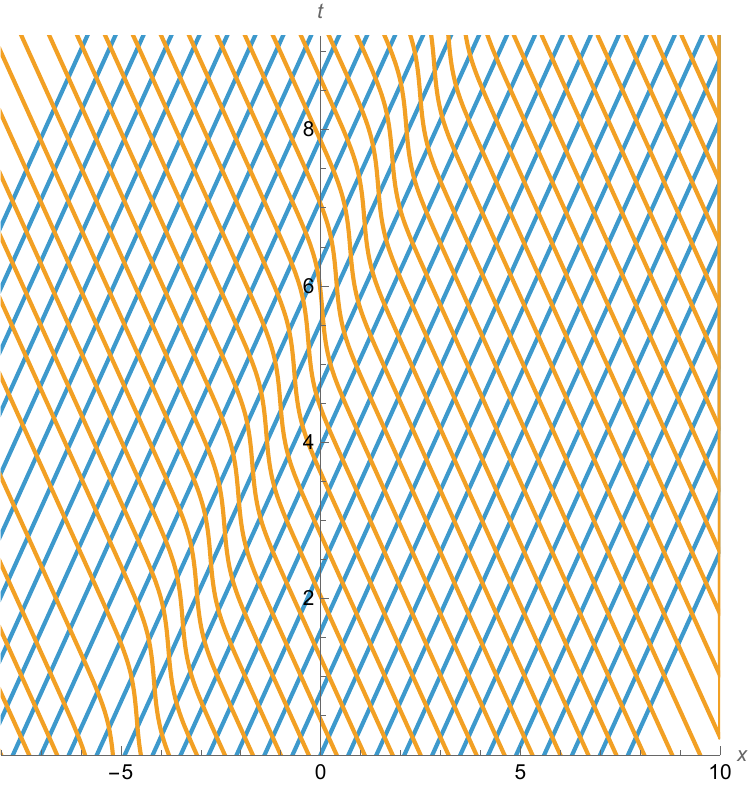} 
        \includegraphics[width=0.3\textwidth]{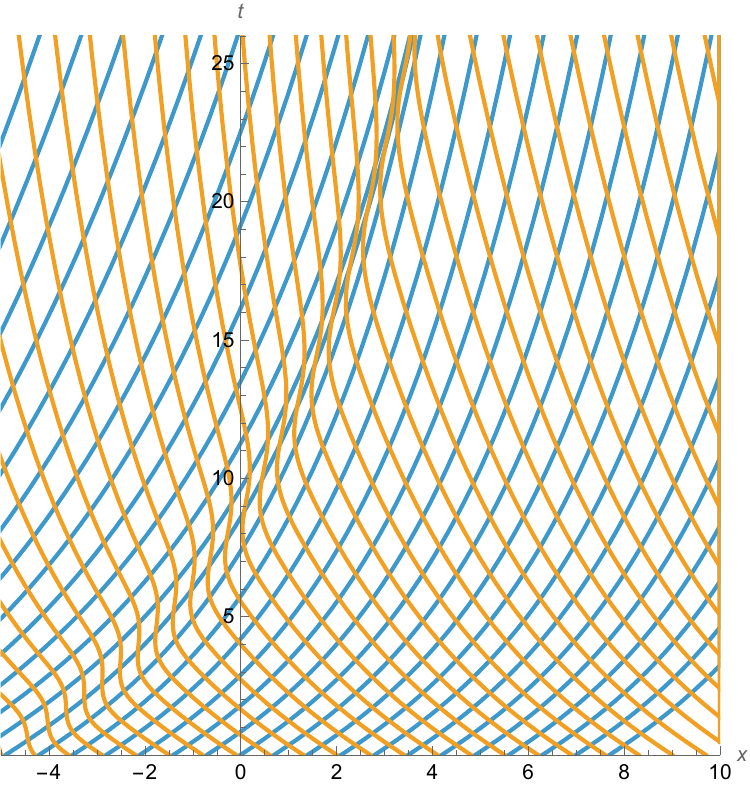} 
            \includegraphics[width=0.3\textwidth]{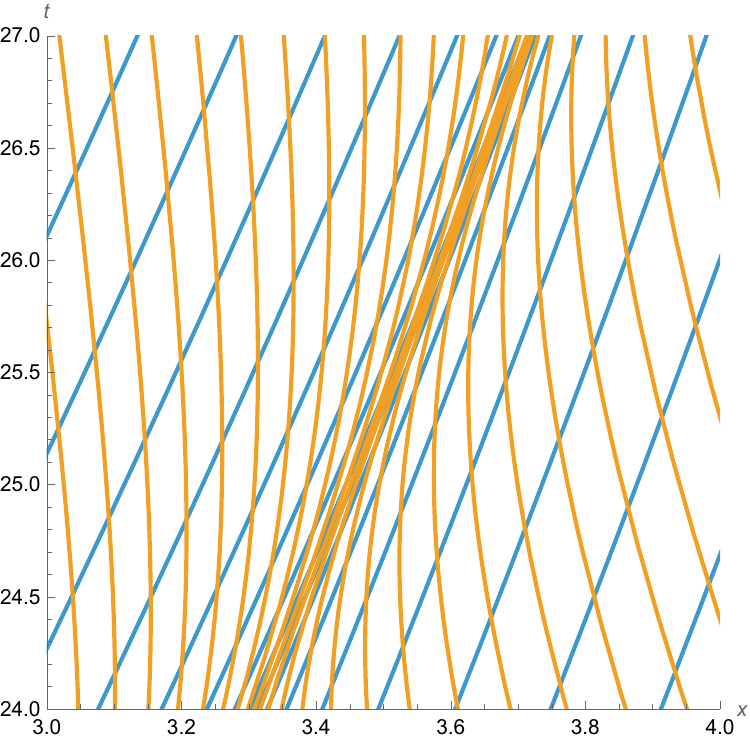} 
\end{center}
    \caption{\textit{Left panel.} Smooth propagation of a wave is demonstrated in the case of DBI in 2D flat spacetime for a particular choice of initial conditions. \textit{Middle panel.} Cusp 
    formation is shown in the case of DBI extended by the $\epsilon$-term described in the end of Sec.~\ref{beyond}, for the same choice of initial conditions as in the left panel. 
    \textit{Right panel.} The same as in the middle panel with a zoom on the region where the cusp is formed.}
     \label{e_DBI_cusp}
\end{figure}

\begin{figure}[!htb]
\begin{center}
    \includegraphics[width=0.3\textwidth]{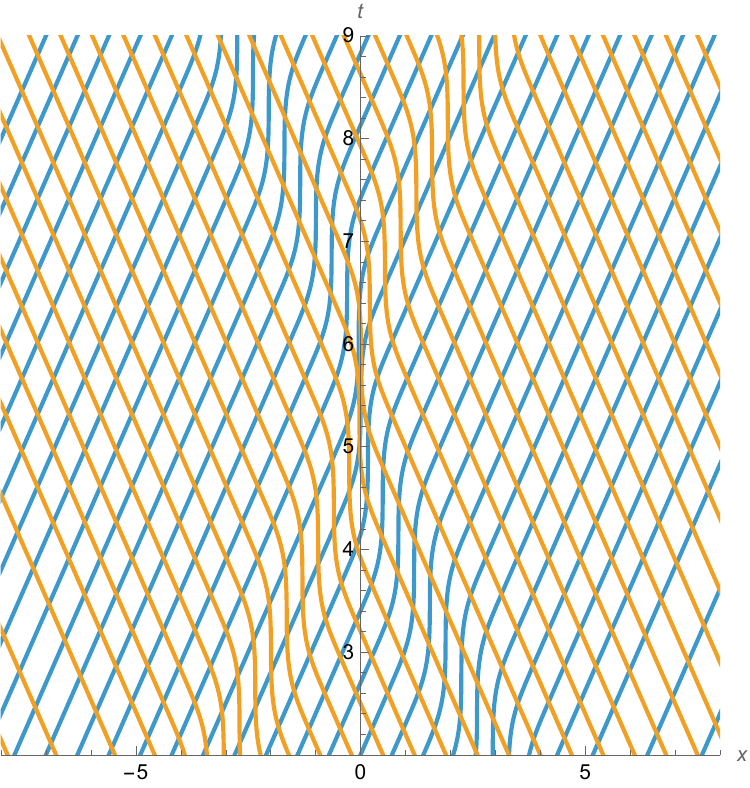} 
        \includegraphics[width=0.3\textwidth]{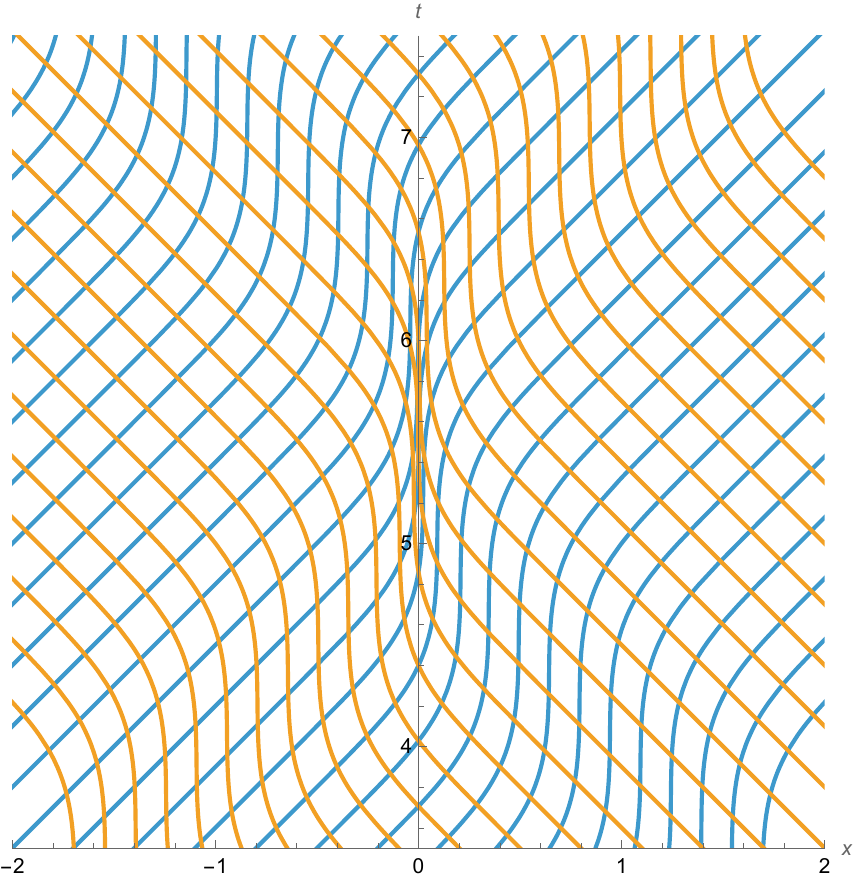} 
            \includegraphics[width=0.3\textwidth]{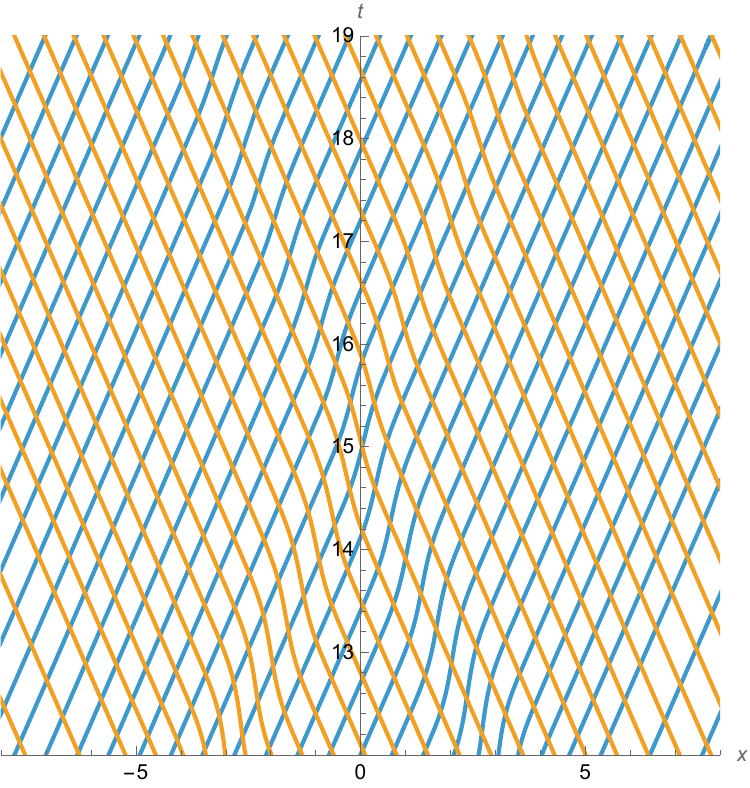} 
\end{center}
    \caption{\textit{Left panel.} Cusp formation is shown in the case of DBI in 2D flat spacetime for a particular choice of initial conditions. \textit{Middle panel.} The same as in the left panel but with a zoom on the cusp region. 
    \textit{Right panel.} Wave propagation is demonstrated in the case of DBI for the same choice of initial conditions as in the left panel, but with cosmic expansion taken into account. No caustics develop in this case.}
     \label{H_DBI}
\end{figure}

Our previous discussion has been focused exclusively on the hyperbolic case, i.e., $\xi_{+} \neq \xi_{-}$, 
which we have proven to be caustic free. 
However, caustics emerge when $\xi_{+}=\xi_{-}$. Indeed, as it is evident from Eq.~\eqref{ddtaut}, the loss of hyperbolicity generally leads to diverging second derivatives of the scalar $\phi$. That singularity, which is cusp-shaped, has been already explored to some extent in the literature in 2D flat spacetime~\cite{Felder:2002sv, Eggers:2009zz, EHS, Blanco-Pillado:2025gzs}. We consider a tractable example of cusp formation in this simplified setup below; details of calculations can be found in Appendix A. We also point out differences of cusp formation in the case of DBI in 2D flat spacetime and beyond, see Figs.~\ref{e_DBI_cusp} and~\ref{H_DBI}. This is in accordance with the discussion in the previous section, where non-trivial patterns of characteristic curves have been uncovered in realistic physical scenarios.

The physical picture behind the cusp becomes clear from Eq.~\eqref{cusper}. One observes that for a finite $\chi$ the loss of hyperbolicity corresponds to $c^2_s=1-2X=0$, or equivalently 
\begin{equation}
\label{press}
\dot{\phi}^2-\phi^{'2}=1 \; , 
\end{equation}
in which case the DBI Lagrangian $P(X)=-\sqrt{1-2X}$ vanishes. Recall that DBI corresponds to the Nambu-Goto action in the static gauge, see Sec.~\ref{todbi}. Hence, vanishing of the DBI Lagrangian means vanishing of the wall area element $d^2 \zeta \sqrt{\gamma}$ for a wall living in 3D (or $d^3\zeta \sqrt{\gamma}$ in 4D). 

\begin{figure}[!htb]
\begin{center}
    \includegraphics[width=0.5\textwidth]{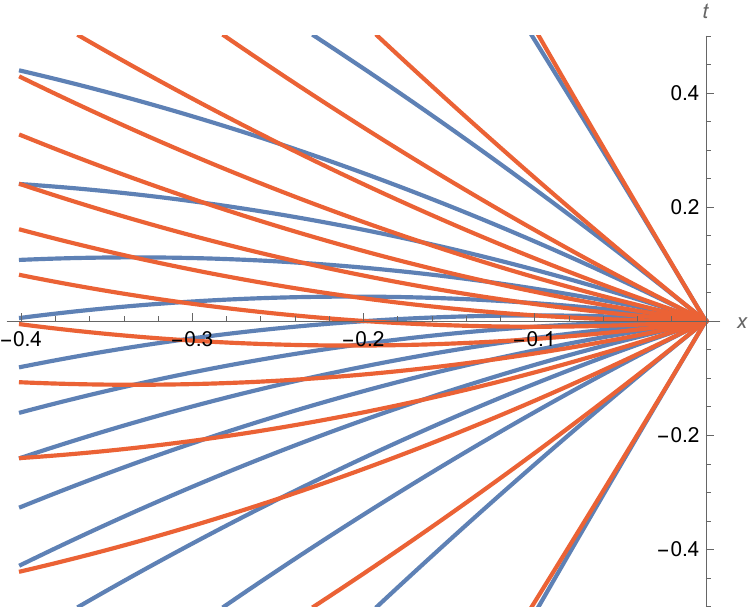} 
\end{center}
    \caption{An example of a ``perfect'' cusp, where all the characteristics from both families converge at one point. 
   See the discussion in Sec.~\ref{cuspy} and in Appendix A. } \label{cusp_c}
\end{figure}

As it follows, in the limit $\xi_{-} \rightarrow \xi_{+}$ DBI can be approximated by a pressureless perfect fluid, and the DBI scalar assumes the role of the velocity potential, i.e., 
$V=-\partial \phi/\partial x$. We again simplify to the case of 2D flat spacetime here, but the singularity for $c_s=0$ emerges beyond this case.
It is well known that the pressureless perfect fluid leads to caustics. Let us show this and meanwhile find a particular solution for the field $\phi$ in the vicinity of a cusp, which we assume to take place at $x=0$ without loss of generality. We perform a formal splitting of the field $\phi$ into a homogeneous part and a perturbation, $\phi=t +\delta \phi (t, x)$. Assuming that $V=-\partial \delta \phi/\partial x$ is small, i.e., $|V| \ll 1$, one gets from Eq.~\eqref{press}:
\begin{equation}
\frac{\partial V}{\partial t}+V\frac{\partial V}{\partial x}=0 \; .
\end{equation}
This is a well-known pressureless Euler equation. If one starts from the initial scalar field profile $V=-x/T$, 
where $T$ is some constant, one can find that the late time solution reads $V=-\frac{x}{T-t}$. Hence, $T$ is the time, when the  spatial derivative of $V$ becomes singular. 
Note that formally $V$ blows up at each $x \neq 0$ when $t \rightarrow T$. It is, however, to be trusted only in the vicinity of $x=0$, 
cf. Ref.~\cite{Babichev:2017lrx}. As a result, at $x \simeq 0$ the solution for the field $\phi$ reads 
\begin{equation}
\label{cusp}
\phi=t+\frac{x^2}{2(T-t)} \; .
\end{equation}
This has a cusp profile for $t \rightarrow T$, as it could be anticipated from the beginning.

In Appendix A, we provide a more consistent derivation of the cusp 
profile starting from the Nambu-Goto action in the conformal gauge. The resulting expression~\eqref{cuspcorr} slightly corrects Eq.~\eqref{cusp}, but the difference does not affect the singularity structure. In particular, Eq.~\eqref{cusp}, where one should take $T=0$, and Eq.~\eqref{cuspcorr} yield the same $\phi''=\chi'$, which is the only second derivative of $\phi$ diverging at the cusp. The fact that $\partial \tau/\partial t$ and $\partial \chi/\partial t$ do not diverge is a particular feature of the example considered above, where all the characteristic lines fulfill $\xi_{+}=\xi_{-}=0$ at $x=0$. This can be seen from the expression:
\begin{equation}
\label{ddchix}
\frac{\partial \tau}{\partial t}=\frac{1}{\xi_{-}-\xi_{+}} \cdot \left[\frac{\partial \tau}{\partial \omega_{+}} \cdot \frac{\xi_{-}}{\partial t/\partial \omega_{+}}-\frac{\partial \tau}{\partial \omega_{-}} \cdot \frac{\xi_{+}}{\partial t/\partial \omega_{-}} \right] \; ,
\end{equation}
which is derived similarly to Eq.~\eqref{ddchix}. The analogous expression holds for $\partial \chi/\partial t$. Clearly, with $\xi_{+}=\xi_{-}=0$ one can avoid a singularity in Eq.~\eqref{ddchix}. 
In Appendix A, we also identify characteristic curves, which are shown in Fig.~\ref{cusp_c}. 
Interestingly, in this particular example all the characteristics converge to one point corresponding to the cusp.

In Sec.~\ref{beyond}, we have pointed out a qualitative difference of wave propagation in DBI in 2D flat spacetime and in more realistic scenarios. These differences have important implications for cusp formation. In particular, in Fig.~\ref{e_DBI_cusp} we consider initial conditions such that no cusps form in DBI in 2D Minkowski space. At the same time, the middle and right panels of Fig.~\ref{e_DBI_cusp} clearly exhibit caustics in the case of DBI extended by the $\epsilon$-term for the same initial conditions. Recall that such a modification discussed in the end of Sec.~\ref{beyond}, is 
invoked for the description of (constant tension) domain walls. The opposite situation is also possible, and it is demonstrated in Fig.~\ref{H_DBI}. Namely, we start with initial conditions such that caustics form in the case of DBI in Minkowski space, and demonstrate that no caustics develop upon switching on cosmic expansion. 
That result is intuitively clear: Hubble friction serves to damp a wave amplitude and thus counteracts cusp formation. 
To summarize Figs.~\ref{e_DBI_cusp} and~\ref{H_DBI}, cusp formation occurs distinctly in DBI in 2D flat spacetime, which is commonly assumed for the sake of simplicity, and beyond.

\section{Discussion}
\label{discussions}
In this work we have studied DBI on the issue of caustic singularities. DBI characteristics have been shown to fulfill the property $\xi_{+}=\xi_{+} (\omega_{+})$ and $\xi_{-}=\xi_{-} (\omega_{-})$ in 2D flat spacetime, which warrants smooth propagation of generic waves in the hyperbolic case. We would like to stress that this property is exclusive for 2D flat spacetime. Namely, it is violated in any realistic cosmological setup, where one has $\partial \xi_{\pm}/\partial \omega_{\mp} \neq 0$ meaning that characteristics are not parallel to each other globally, see Sec.~\ref{beyond}. Nevertheless, DBI exhibits a remarkable resistance to caustics. 
In particular, caustics do not form in the hyperbolic case, if the principal part of the PDE for the DBI field has the same form as in the 2D flat spacetime. This holds true in  $D>2$ in the tractable case of spherical waves, as well as in $D=2$ but with the Universe expansion turned on. It is straightforward to generalize this conclusion to an arbitrary curved background metric in $D=2$. The question of DBI field propagation in more than two dimensions beyond spherical symmetry remains open, and we leave it for future work.

This strongly indicates that DBI caustics, whenever observed in simulations, are directly associated with the loss of hyperbolicity, or put differently, to vanishing of the sound speed $c_s$. So far, numerical analysis of caustic formation in DBI has been limited to flat 2D spacetime~\cite{Felder:2002sv, Blanco-Pillado:2025gzs}. 
However, it is clear from Figs.~\ref{e_DBI_cusp} and~\ref{H_DBI} that the manifestation of caustics in the non-hyperbolic case (cusps), in particular the possibility of their formation or avoidance, is significantly affected beyond this simplified setting.  
The physical reason is the aforementioned drastic difference of characteristic curve patterns in DBI in 2D flat spacetime 
and beyond. Therefore, one should be cautious when extrapolating 
conclusions from numerical simulations based on the simplified action~\eqref{dbi} to realistic scenarios described by Eq.~\eqref{dbireal} or Eq.~\eqref{dbilinear}.

The search for caustics undertaken in this work has been motivated by their potentially important role for the particle production by domain walls\footnote{Another mechanism of particle 
emission from a domain wall based on parametric resonance has been discussed in Ref.~\cite{Blanco-Pillado:2022rad}.}. As it follows, the loss of hyperbolicity is likely to be the main reason for the particle production. Let us briefly speculate on the mechanism of particle emission via formation of non-hyperbolic caustics, leaving a detailed study of this question for future work. 
Recall that the effective description of a planar domain wall in terms of DBI assumes single-valuedness of the field $\phi \equiv z$ and its derivatives. Cusp formation predicted by DBI exactly means that this assumption cannot be maintained at all the times, and hence the wall effective description should necessarily break down at some point. 
Namely, DBI effective description fails to capture the realistic dynamics of domain walls in near-cusp regions, and one should eventually resort to UV description provided by Eq.~\eqref{dw} or~\eqref{eqn source pi}. In this UV complete picture existence of a heavy degree of freedom associated with the field $\Psi$ is manifest, and this degree of freedom is naturally excited where cusps are formed in the DBI description.  The breakdown of the effective description can occur in either of two ways. i) Finite width effects become relevant, and consequently the Nambu-Goto action (and hence the DBI action) ceases to correctly describe the dynamics of the domain wall, in particular due to the excitation of heavy particles associated with the field constituting the wall. ii) The DBI field description breaks down while the Nambu-Goto approximation still holds. In this case, the field $\phi$ tends to become multi-valued, violating the assumption underlying the DBI description\footnote{This is reminiscent of the situation with pressureless perfect fluid, where caustics manifest the onset of multistream regions.}. In that case, one expects formation of closed walls detaching from a long wall. These closed walls eventually shrink and produce particles of the field comprising the domain wall~\cite{Widrow:1989vj}. In either case, we expect particle production to be the result of caustic formation.

Note also that there are different types of domain walls, and each type may have a distinct caustic pattern. For example, the so called melting domain walls~\cite{Ramazanov:2021eya, Babichev:2021uvl}, characterized by a time-varying tension, effectively live in Minkowski space and do not require $\epsilon$-term, since their existence does not pose problems in cosmology. 
Thus, the structure of their characteristics can be significantly different from that of conventional constant tension domain walls assumed in Sec.~\ref{beyond}. It would be interesting to investigate, if such differences can be responsible for the apparent dissimilarities seen
in numerical analysis of conventional and melting domain walls~\cite{Dankovsky:2024ipq}.

As a final remark, it is worth pointing out broader theoretical implications of our results, beyond domain walls.
Namely, modified gravity theories are often prone to caustic formation~\cite{Babichev:2016hys} and other pathologies, such as the emergence of ghosts~\cite{Babichev:2020tct}. In this context one naturally deals with two types of caustics: those caused by the loss of hyperbolicity, as in the DBI case, and those, which arise in the hyperbolic case~\cite{Babichev:2016hys}. It would be interesting to study possible differences in manifestations of these two types of caustics, and to investigate which ones are more likely to form in a cosmological setup. In this broader context, it is also worth studying phenomenological consequences of caustics, e.g., emission of extra degrees of freedom arising in the high energy completion of corresponding effective theories~\cite{Babichev:2017lrx,Babichev:2018twg}. Another natural direction of future research is the investigation of caustics in superluminal DBI theories. We leave these questions for forthcoming studies.

\section*{Acknowledgments}
The work of EB was supported by ANR grant StronG (No. ANR-22-CE31-0015-01). The work of MVV was supported by Russian Science Foundation grant No. 24-72-10110,\\https://rscf.ru/project/24-72-10110/.

\section*{Appendix A. Cusp in conformal and static gauges}

As it has been discussed in Sec.~\ref{todbi}, DBI corresponds to the Nambu-Goto action~\eqref{NG} in the static gauge. 
In this appendix we consider the conformal gauge, which is also widely discussed in the literature~\cite{Vilenkin}. The conformal gauge corresponds to the choice $\zeta^0=t$ and the condition on the worldsheet metric $\gamma_{ab}=\sqrt{|\gamma|} \eta_{ab}$. We use the notation $\theta$ for the second coordinate on the wall, i.e., $\zeta^2 \equiv \theta$. 
As in the main text, we restrict to domain walls in 3D. With the choice $\zeta^{0}=t$, the induced metric on the planar wall living in 3D can be written as
\begin{equation}
\gamma_{a b}=\left(\begin{array}{cc}|\vec{X}_t|^2-1 & \vec{X}_t \vec{X}_{\theta} \\ \vec{X}_t \vec{X}_{\theta} & \left|\vec{X}_{\theta}\right|^2\end{array}\right) \; ,
\end{equation}
where $\vec{X} \equiv (x, z)$. Here we use the subscripts $t$ and $\theta$ to denote the corresponding partial derivatives. The former is different from the dot notation mainly used in the text; in this way we differentiate partial time derivatives at fixed $x$ and at fixed $\theta$ (when the conformal gauge is concerned). The condition for the metric to be conformal reads
\begin{equation}
\label{cons}
\left\{\begin{array}{l}
\vec{X}_t \vec{X}_{\theta}=0 \\
\left|\vec{X}_t \right|^2+\left|\vec{X}_{\theta}\right|^2=1 \; .
\end{array}\right.
\end{equation}
In the conformal gauge, evolution of the vector $\vec{X}$ following from the Nambu-Goto action is simply described by the wave equation:
\begin{equation}
\eta^{a b} \partial_a \partial_b \vec{X}=0 \; .
\end{equation}
The solution of this equation can be written as
\begin{equation}
\label{waveeq}
\vec{X}(t, \theta)=\frac{1}{2}[\vec{a}(\theta-t)+\vec{b}(\theta+t)] \; ,
\end{equation}
and the gauge fixing condition takes the form:
\begin{equation}
\label{constra}
|\vec{a}_{\theta}|=|\vec{b}_{\theta}|=1 \; .
\end{equation}

Recall that in DBI corresponding to the static gauge, one identifies $z$ with the scalar field $\phi$, i.e., $z \equiv \phi$. 
Then, we observe that an arbitrary function $F(t, \theta)$ satisfies 
\begin{equation}
\label{dergauge}
\begin{aligned}
& \dot{F} \equiv \left.\frac{\partial F}{\partial t}\right|_x=F_t+F_\theta \theta_t=F_t-\frac{x_t}{x_\theta} F_\theta \\
& F' \equiv \left.\frac{\partial F}{\partial x}\right|_t=F_\theta \theta_x=\frac{1}{x_\theta} F_\theta \; .
\end{aligned}
\end{equation}
These can be used to calculate the first and the second derivatives of $\phi$, and verify that the equation of motion for the DBI field
\begin{equation}
\left(1+\phi^{\prime 2}\right) \ddot{\phi}-2 \dot{\phi} \phi^{\prime} \cdot \dot{\phi}^{\prime}-\left(1-\dot{\phi}^2\right) \phi^{\prime \prime}=0 \; ,
\end{equation}
is satisfied~\cite{Blanco-Pillado:2025gzs}. This serves as a consistency check between two different gauges.

Now let us find the cusp profile using the conformal gauge and justify Fig.~\ref{cusp_c}. In the conformal gauge, the cusp is defined as ${ \vec { X }_t  = 1 }$~\cite{Vilenkin}, which translates into $\vec{X}_{\theta}=0$ by virtue of Eq.~\eqref{cons}, so that
$\vec{a}_{\theta}=-\vec{b}_{\theta}$ at the cusp. We are free to choose $\vec{a}_{\theta}=-\vec{b}_{\theta}=\{0,-1\}$, which corresponds to $\dot{\vec{X}}$ lying along $z$-axis. Without loss of generality, we assume that the cusp takes place at $t=\theta=0$.
Then, keeping only the first non-trivial terms in the expansion of $\vec{X}$ in the vicinity of the cusp, one writes
\begin{equation}
\begin{aligned}
&x (t,\theta)=\frac{1}{4}\left(\alpha(t-\theta)^2+\beta(t+\theta)^2\right) \\
&z (t, \theta)=t-\frac{1}{12}\left(\alpha^2(t-\theta)^3+\beta^2(t+\theta)^3\right)  \; ,
\end{aligned}
\end{equation}
where $\alpha$ and $\beta$ are some constants. Here we have used the constraint~\eqref{constra}. In the so called bifurcation case~\cite{Blanco-Pillado:2025gzs}, i.e., $\alpha+\beta=0$, the above expressions take the form:
\begin{equation}
\label{xz}
\begin{aligned}
& x(t, \theta)=-\alpha t \theta \\
& z (t, \theta)=t-\frac{\alpha^2}{12}\left((t+\theta)^3+(t-\theta)^3\right) \; .
\end{aligned}
\end{equation}
Using the expansion for $x,z$ and Eq.~(\ref{dergauge}), we can switch to the static gauge and find 
the derivatives of the DBI field:
\begin{equation}
\label{above}
\begin{aligned}
& \tau=z_t-\frac{\dot{x}}{x_\theta} z_\theta=1-\frac{\alpha^2}{2}\left(t^2-\theta^2\right)  \\
& \chi=\frac{z_{\theta}}{x_{\theta}}= \alpha \theta \; .
\end{aligned}
\end{equation}
As a result, the DBI sound speed squared can be written as $c_s=\sqrt{1-\tau^2+\chi^2}=|\alpha t|$ near the cusp. Substituting $c_s=|\alpha t|$ and Eq.~\eqref{above} into Eq.~\eqref{csv}, one obtains the slopes of characteristics near the cusp:
\begin{equation}
\label{xicusp}
\xi_{\pm} = \pm |\alpha t|-\alpha \theta \; .
\end{equation}
Now we can find characteristic curves in the $(t, x)$ plane. Using Eqs.~\eqref{xz},~\eqref{xicusp} and
\begin{equation}
\frac{d x(t, \theta(t))}{d t}=x_t+x_\theta \frac{d \theta}{d t}=\xi_{ \pm}(t, \theta) \; ,
\end{equation}
we obtain
\begin{equation}
-\left.\alpha t \frac{d \theta}{d t}\right|_{\omega_{ \pm}}= \pm|\alpha t| \; .
\end{equation}
Taking $t<0$, since the cusp is just forming, and $\alpha>0$ without loss of generality, we get
\begin{equation}
\left.\theta\right|_{\omega_{+}}=d_{+}+t \quad \left.\theta\right|_{\omega_{-}}=d_{-}-t \; ,
\end{equation}
where $d_{\pm}$ are some constants. Our choice $t=\theta=0$ at the cusp enforces $d_{\pm}=0$. Taking into account $x=-\alpha t \theta$, we can write
\begin{equation}
\left.x\right|_{\omega_{ \pm}}= \mp \alpha t^2.
\end{equation}
We observe that all characteristics meet at one point of the cusp singularity. Furthermore, for some characteristics there are turning points, where they change direction from right to left and vice versa for the other family.

Finally, we can use Eq.~\eqref{xz} to express 
the DBI field $\phi \equiv z$ in terms of $t$ and $x$: 
\begin{equation}
\label{cuspcorr}
\phi =t -\frac{\alpha^2 t^3}{6} -\frac{x^2}{2t} \; .
\end{equation}
This corrects Eq.~\eqref{cusp}, which has been obtained assuming $|\phi'| \ll 1$ and $c_s=0$ in the vicinity of the cusp.

\section*{Appendix B. Lax approach to caustic formation}

Let us discuss Lax approach for the study of caustic formation assuming that PDE~\eqref{partial} is strictly hyperbolic~\cite{lax1964development, lax1973hyperbolic, lax1955xii}. We start with Eq.~\eqref{slopes}.
 For concreteness, let us focus on the first of Eq.~\eqref{slopes} and apply the spatial derivative to it. Introducing a new variable $u \equiv \frac{\partial \omega_{+}}{\partial x}$, we obtain
\begin{align}
        \frac{\partial u}{\partial t } +\xi_-\;\frac{\partial u}{\partial x} +\frac{\partial \xi_-}{\partial \omega_+}u^2+\frac{\partial \xi_-}{\partial \omega_-}\;\frac{\partial \omega_-}{\partial x }\;u &=0\;,\label{eqn u1}
\end{align}
which is a Riccati equation for $u$. In what follows it is convenient to introduce the derivatives along the $\xi_{+}$ and $\xi_{-}$-characteristics: 
\begin{equation}
   D_+=\frac{\partial}{\partial t}+\xi_+\;\frac{\partial}{\partial x} \qquad    D_-=\frac{\partial}{\partial t}+\xi_-\;\frac{\partial}{\partial x}\;.\label{eqn D+}
\end{equation}
Note that Eq.~\eqref{slopes} takes a particularly simple form in terms of these derivatives: 
\begin{equation}
\label{simp}
D_{-} \omega_{+}=0 \qquad D_{+} \omega_{-}=0 \; .
\end{equation}
Using Eq.~\eqref{eqn D+}, one can rewrite Eq.~(\ref{eqn u1}) as an ordinary differential equation for $u$ along $\xi_-$-characteristics,
\begin{align}
        D_-u +\frac{\partial \xi_-}{\partial \omega_+}u^2+\frac{\partial \xi_-}{\partial \omega_-}\;\frac{D_-\omega_-}{(\xi_- -\xi_+)}\;u=0\;.\label{eqn u2}
\end{align}
Here we have also used Eqs.~\eqref{eqn D+} and~\eqref{simp} to express derivatives of $\omega_{\pm}$ with respect to $t$ and $x$ in terms of $D_{\pm}$ derivatives, in particular 
\begin{equation}
\frac{\partial \omega_{-}}{\partial x}=\frac{D_{-} \omega_{-}}{\xi_{-}- \xi_{+}} \; .
\end{equation}
Now defining
$D_-h \equiv \frac{\partial \xi_-}{\partial \omega_-}\;\frac{D_-\omega_-}{(\xi_- -\xi_+)}$, we rewrite it as
\begin{align}
        D_-(e^h\;u)+e^{-h}\;\frac{\partial \xi_-}{\partial \omega_+}(e^{h}\;u)^2=0\;.\nonumber
\end{align}
Defining  $q(t)=e^h\;u=e^h\;\frac{\partial \omega_+}{\partial x}\;,$ we obtain
\begin{align}
   D_-\;q+e^{-h}\;\frac{\partial \xi_-}{\partial \omega_+}\;q^2=0\;.\label{eqn ufinal0}
\end{align}
Recognizing the derivative $D_{-}$ as the full time derivative along the $\xi_{-}$-curve, i.e., $D_-=\frac{d}{dt}\Big\vert_{x^\mu}$, one writes the solution of Eq.~\eqref{eqn ufinal0} as follows:
\begin{align}
    q(t)=\frac{q_0}{1+q_0\;\int^t_{t_0} dt'\Big(e^{-h}\;\frac{\partial \xi_-}{\partial \omega_+}\Big)}\;,\label{eqn ufinal}
\end{align}
where $q_0$ is the initial condition for $q(t)$. Formation of caustics is linked to vanishing of the denominator in Eq.~\eqref{eqn ufinal}, at which point $q$ explodes, because $q \propto \partial \omega_{+}/\partial x$. 

It is clear that the denominator never vanishes in Eq.~\eqref{eqn ufinal}, if $\partial \xi_{-}/\partial \omega_{+}=0$ identically, 
as it is the case of DBI in 2D flat spacetime. 
Hence, no caustics form in this case. On the other hand, the fact that $\partial \xi_{-}/\partial \omega_{+} \neq 0$ does 
not warrant shock formation, i.e., if one cannot fulfill the condition $|e^{-h} \partial \xi_{+}/\partial \omega_{-}| >\mbox{c}>0$, where $c$ is some fixed constant. In particular, it cannot be fulfilled in DBI beyond 2D flat spacetime, and we have 
indeed seen in Sec.~\ref{beyond} that no caustics develop in this case either.

\section*{Appendix C. Action for thin annihilating walls}

In this Appendix we comment on the derivation of Eq.~\eqref{dbilinear} from Eq.~\eqref{eqn source pi} in the thin wall limit. For this purpose, we have followed the approach of Refs.~\cite{Israel:1966rt,Berezin:1987bc} designed for the study of thin self-gravitating shells; see also Refs.~\cite{Deng:2017uwc, Deng:2020mds}, where this approach is applied to spherical walls. One describes the domain wall in this approach as a thin timelike shell that separates two regions of spacetime denoted by $+$ and $-$. Note that we assume the metric of the expanding Universe here with the line element $ds^2=-dt^2+a^2 (dx^2+dy^2+dz^2)$.
The discussion in Ref.~\cite{Berezin:1987bc} dramatically simplifies, provided that one neglects self-gravity 
of a domain wall, i.e., we consider the formal limit of the Planck mass to infinity, $M_{Pl} \rightarrow \infty$. In this case the metric and the extrinsic 
curvature are the same on both sides of the wall, and the scope of equations in Ref.~\cite{Berezin:1987bc} reduces to a single one\footnote{The equation below matches Eq.~(2.32a) in Ref.~\cite{Berezin:1987bc} upon substituting the shell stress energy tensor $S^{a}_{b}=-\sigma \delta^{a}_{b}$, which is justified for vacuum shells, as it is explained near Eq.~(2.66) there. Furthermore, Eq.~(2.32b) in Ref.~\cite{Berezin:1987bc} is automatically satisfied for this choice of $S^{a}_b$ assuming a constant tension $\sigma$, see Eq.~(2.67).}:
\begin{align}
    K=\frac{1}{\sigma}[T_{\mu\nu}\;\xi^\mu\;\xi^\nu]\, ,\label{eqn eom wall2}
\end{align}
where $T_{\mu \nu}$ is the stress-energy tensor associated with the domain wall field $\Psi$, $\sigma$ is the wall tension, and
\begin{equation}
\label{wrapped}
K \equiv g^{\mu \nu} \nabla_{\mu} \xi_{\nu} \; .
\end{equation}
Here $K$ is the trace of extrinsic curvature tensor given by $K_{\mu\nu}=\nabla_\mu\xi_\nu$, and $\xi^{\mu}$ is a spacelike unit normal
vector to the wall~\cite{Poisson:2009pwt}. The square brackets on the r.h.s. of Eq.~\eqref{eqn eom wall2} mean that one takes 
the difference of a corresponding quantity on two sides of the wall, i.e., 
\begin{equation}
[T_{\mu\nu}\;\xi^\mu\;\xi^\nu] \equiv  {}^+(T_{\mu\nu}\;\xi^\mu\;\xi^\nu)-{}^-(T_{\mu\nu}\;\xi^\mu\;\xi^\nu )\; .
\end{equation}
The stress-energy tensor $T_{\mu \nu}$ is inferred from the action~\eqref{eqn source pi}. With no loss of generality, one writes for the domain wall field $\Psi$ on two sides of the wall ${}^\pm\Psi(x)=\pm \langle \Psi \rangle+{}^{\pm}\psi(x)$, where ${}^{\pm}\psi$ are fluctuations, which we ignore in what follows. Namely, we are focusing solely on a vacuum configuration of the field $\Psi$. Noting that ${}^\pm g_{\mu\nu}=g_{\mu\nu}$ and $g_{\mu\nu}\xi^\mu\xi^\nu=1$, we obtain from Eq.~(\ref{eqn source pi}) that 
\begin{equation}
\label{con}
[T_{\mu\nu}\;\xi^\mu\;\xi^\nu]=\epsilon. 
\end{equation}
Recall that $\epsilon$ is a constant responsible for the explicit symmetry breaking in Eq.~\eqref{eqn source pi} and consequently 
for the domain wall annihilation. It is crucial that the exponent $m$ is odd in Eq.~\eqref{eqn source pi}, --- otherwise, 
there would be no breaking of $Z_2$-symmetry, and the r.h.s. of Eq.~\eqref{con} would vanish.

To calculate the l.h.s. of Eq.~\eqref{eqn eom wall2}, we need to know the normal vector $\xi^{\mu}$. For this purpose we have 
to define the relation between the coordinates $\zeta^{a}$, $a=0,1,2$, on the wall and the 4D spacetime coordinates $x^{\mu} =(t, x, y, z)$. We choose $\zeta^{0}=t$, $\zeta^{1}=x$, and $\zeta^{2}=y$, while the third spatial coordinate $z$ is identified with the scalar $\phi$, cf. Sec.~\ref{todbi}. 
The normal vector is then fixed by the conditions $\xi_{\mu} \partial_a x^{\mu}=0$ and $g_{\mu \nu} \xi^{\mu} \xi^{\nu}=1$:
\begin{align}
    \xi^\mu=\frac{1}{\sqrt{1-2X}}\left(a\dot{\phi},\, -\frac{\phi_x}{a},\, -\frac{\phi_y}{a},\,\frac{1}{a}\right)\, ,\label{eqn normal}
\end{align}
cf. Ref.~\cite{Garriga:1991ts}. Here the subscripts $x$ and $y$ denote partial derivatives with respect to $x$ and $y$, and $X=(a^2\dot{\phi}^2 -\phi^2_x-\phi^2_y)/2$, which matches Eq.~\eqref{X}. Having calculated $\xi^{\mu}$, one can obtain $K$ with the use of Eq.~\eqref{wrapped}. Substituting the result for $K$ into Eq. (\ref{eqn eom wall2}), using Eq.~\eqref{con}, and redefining the coordinates $x^{\mu}=(\eta, x, y)$, where $\eta$ is the conformal time, we get
\begin{align}
    -\eta^{\mu \nu} \frac{\partial}{\partial x^{\mu}} \left(\frac{1}{\sqrt{1-2X}} \cdot \frac{\partial \phi}{\partial x^{\nu}} \right)+\frac{3{\cal H}}{\sqrt{1-2X}} \frac{\partial \phi}{\partial \eta} =\frac{\epsilon a}{\sigma}\;,\label{eqn biased DBI}
\end{align}
where $\eta^{\mu \nu}$ is the 3D Minkowski metric and ${\cal H}=\partial \ln a/\partial \eta$. One recognizes Eq.~\eqref{eqn biased DBI} as the equation of motion for the DBI field $\phi$ following from Eq.~\eqref{dbilinear}.

\end{document}